\documentclass[onecolumn]{elsart3p}        

\usepackage{amsmath,amssymb,epsfig}
\usepackage{graphics}
\usepackage{changebar}
\usepackage{graphicx}
\usepackage{graphics}
\usepackage[dvips]{color}
\usepackage{bm}
\usepackage[square,comma]{natbib}
\journal{Journal of Computational Physics}
\psfigdriver{dvips}

\renewcommand{\vec}[1]{{\bf #1}}
\newcommand{\mat}[1]{{\bf #1}}   

\def\secref#1{section~\ref{#1}}

\def\fig#1{Figure~\ref{#1}}
\def\figs#1#2{Figures~(\ref{#1}--\ref{#2})}

\def\eqref#1{Eq.~(\ref{#1})}
\def\Eq.#1{Eq.~(\ref{#1})}
\def\Eqs.#1#2{Eqs.~(\ref{#1}--\ref{#2})} 
\def\Eq#1{Equation~(\ref{#1})}
\def\Eqs#1#2{Equations~(\ref{#1}--\ref{#2})} 
\def\eq#1{equation~(\ref{#1})}
\def\eqs#1#2{equations~(\ref{#1}--\ref{#2})}

\def\bea{\begin{eqnarray}}
\def\eea{\end{eqnarray}}

\def\be{\begin{equation}}
\def\ee{\end{equation}}

\def\Bperp{\bm B_{\perp}}
\def\ez{\bm{e}_{z}}

\def\kperp{k_\perp}
\def\nabperp{\nabla_\perp}
\def\roi{\rho_{i}}
\def\ros{\rho_s}
\def\rhotau{\rho_\tau}
\def\ne1{n_{e1}}

\def\({\left(}
\def\){\right)}
\def\[{\left[}
\def\]{\right]}
\def\<{\left\langle}
\def\>{\right\rangle}

\def\DD{\mathcal D}
\def\LL{\mathcal L}
\def\GG{\mathcal G}
\def\deta{\mathcal D_\eta}
\def\dnu{\mathcal D_\nu}
\def\d{\partial}
\def\D'{\Delta'}

\def\roi{\rho_{i}}
\def\etal{\textit{et al.}}
\def\dt{\Delta t}

\def\Error{\mathcal E}
\def\Err_max{\mathcal E^{max}}
\def\pmax{p_{max}}
\def\FF{\mathcal F}
\def\etak{\deta}
\def\nuk{\dnu}

\begin{document}
\begin{frontmatter}


 \title{An iterative semi-implicit scheme with robust damping}

 \author[CMPD,PPPL]{N.~F.~Loureiro\corauthref{cor}}
 \ead{loureiro@pppl.gov}
 \ead[url]{http://www.cscamm.umd.edu/cmpd/people/loureiro/}
 \corauth[cor]{Corresponding author}
 \author[PPPL]{G.~W.~Hammett}
 \ead{hammett@pppl.gov}
 \ead[url]{http://w3.pppl.gov/$\sim$hammett/}
 \address[CMPD]{CMPD, University of Maryland, College Park, MD 20742, USA}
 \address[PPPL]{Princeton Plasma Physics Laboratory, Princeton University, Princeton, NJ 08543, USA}

\begin{abstract}
An efficient, iterative semi-implicit (SI) numerical method for the time
integration of stiff wave systems is presented. Physics-based assumptions
are used to derive a convergent iterative formulation of the SI scheme
which enables the monitoring and control of the error introduced by the SI
operator.   This iteration essentially turns a semi-implicit method into a
fully implicit method.  Accuracy, rather than stability, determines the
timestep. The scheme is second-order accurate and shown to be equivalent to
a simple preconditioning method. 
We show how the diffusion operators can be handled so 
as to yield the
property of robust damping, i.e., dissipating the solution at all
values of the parameter $\mathcal D\dt$, where $\mathcal D$ is a
diffusion operator and $\dt$ the timestep. The
overall scheme remains second-order accurate even if the advection and
diffusion operators do not commute.
In the limit of no physical dissipation, and for a linear test wave problem, the method is shown to be symplectic.
The method is tested on the problem of Kinetic Alfv\'en wave mediated
magnetic reconnection. A Fourier (pseudo-spectral) representation is used. A 2-field gyrofluid model is used and an efficacious
k-space SI operator for this problem is demonstrated.  CPU speed-up factors over
a CFL-limited explicit algorithm ranging from $\sim20$ to several hundreds
are obtained, while accurately capturing the results of an explicit
integration. 
Possible extension of these results to a real-space (grid) discretization
is discussed.
\end{abstract}

\begin{keyword}
Semi-implicit methods\sep robust damping\sep magnetic reconnection\sep gyrofluid\sep dispersive plasma waves \sep symplectic methods

\PACS 52.30Ex\sep 52.35Vd\sep 52.35Py\sep 52.35Hr\sep 52.65Tt
\end{keyword}
\end{frontmatter}

\section {Introduction}

One ubiquitous property of the dynamics of fluids and plasmas is their
inherent multiscale character, both spatial and
temporal~\cite{scales2,tang_05}. This translates into a stiffness of the
partial differential equations (PDEs) that are used to describe the
physical systems of interest. The numerical solution of stiff systems of
coupled PDEs is  notoriously difficult, presenting a serious challenge to
present day computational resources.

%

Wave-induced temporal stiffness occurs when the characteristic time scale of interest is much slower than that of the fastest waves present in the system. 
Fortunately, however, it is often the case
that it is not necessary to accurately capture these waves in order to obtain accurate solutions of the slowly evolving physics~\cite{boyd, durran}. 
Examples are the gravity wave in weather modeling~\cite{durran,andre_85}, or the Alfv\'en and whistler waves in studies of some instabilities in plasmas~\cite{schnack_87,harned_mikic_89,glasser_99,sovinec_04,chacon_02,chacon_03,reynolds_06}.
When that occurs, 
implicit and semi-implicit (SI) numerical methods have been employed as
very powerful tools to potentially allow the stable and accurate time
integration of the PDEs on timescales which largely exceed those imposed by
these high frequency, fast waves (see, for example, the paper by Knoll
$\etal$~\cite{knoll_05} and references therein). More conventional and easy
to implement explicit numerical schemes are limited by the well-known
Courant-Friedrichs-Lewy (CFL) condition and thus, in order to be stable,
require integration timesteps of the order of the inverse of the maximum
frequency present -- a very severe, and generally unfeasible, requirement.

There are advantages and disadvantages to both purely implicit and
SI numerical methods. 
Implicit and SI methods can both use time steps much larger than the CFL
limit. Implicit methods are more robust and accurate but can be harder to
implement, in some cases requiring more extensive modifications to the
structure of an existing code.  Direct solution of an implicit scheme can
require the storage and inversion of impractically large matrices (though
direct sparse matrix solvers with reordering can be effective for some
problems that are not too large).  
Thus implicit methods often employ preconditioned iterative methods (such
as Krylov methods like Conjugate-Gradients or GMRES), sometimes in
combination with multigrid or FFT solvers (see for example references
\cite{knoll_keyes_04,knoll_05}, which also discuss nonlinear problems using
Jacobian-Free Newton-Krylov solvers).  To obtain a net 
speed up, it is often crucial
to have a sufficiently good preconditioner that can still be efficiently
inverted.  While there are many options
\cite{knoll_keyes_04,knoll_05,petsc-manual,PETSC}, finding such a
preconditioner is problem-dependent and can be a non-trivial task.  If no
preconditioner is used, a Krylov iterative solver applied directly to a
hyperbolic wave/advection problem will usually give at most an order unity
speed up in net CPU time, because the larger time step allowed by an
implicit method is largely offset by the increased number of iterations
required per time step.  However, when a good preconditioner exists, or a
multigrid or FFT solver can be used, then significant savings in CPU time
are possible.

SI methods are usually simpler to implement and work well for some
problems.  However, they have to be used with some care, because in some
parameter regimes there can be splitting errors (due to the different
treatment of the operators involved) and/or uncontrolled approximations to
the equations that are made in their derivation (such as linearization of
the fast wave physics).  
The reader is referred to the work of Knoll and Keyes~\cite{knoll_keyes_04} and Knoll $\etal$~\cite{knoll_05} for an overview of SI methods in
a variety of applications and their comparison to the JFNK approach to
fully implicit methods .

In this paper, we revisit SI methods as they are traditionally
formulated in the field of plasma
physics~\cite{harned_kerner_85,harned_86,schnack_87,caramana_91}. Using
physics-based
assumptions, we are able to consistently derive an iterative formulation of
SI schemes, which lends itself to an efficient way of monitoring and
controlling the error introduced by the SI operator. Accuracy, rather than
stability, determines the timestep.  It is shown that this iterative
extension of an SI scheme leads to essentially a fully implicit method with
the SI operator providing a preconditioner.
We also demonstrate how the diffusive terms can
be treated so as to yield the property of robust damping, i.e., the
consistent damping of all the modes present in the simulation domain in the
limit $\dt \rightarrow \infty$. The SI method presented here is
second-order accurate even if the diffusion and nonlinear advection
operators do not commute.

The method is then applied to the problem of gyrofluid magnetic
reconnection. This is a very stiff application due to the presence of a
high frequency dispersive wave, the Kinetic Alfv\'en wave (KAW). A 
Fourier-space, pseudo-spectral, discretization of the equations is used.
Compared to explicit integrations schemes, our SI scheme yields CPU speed-up factors
of $\sim20$ to several hundreds while accurately reproducing the results of
an explicit integration. 

Before proceeding further, we caution the reader that, 
although our derivation of the SI algorithm presented in 
\secref{derivation_method} is 
general, we have only tested it in a pseudo-spectral implementation, and
only for a 2-field problem that involves a single wave family (a single
dispersion relation $\pm \, \omega$ vs.~$\vec k$).
The efficiency of this method relies heavily on the effectiveness of the SI
operator used, which may be complicated for problems with strong spatial
inhomogeneities, strong anisotropies, or for multi-field problems with
multiple waves.
Also, our treatment of diffusion terms with a robust damping algorithm has
been tested only in 
a Fourier representation, although we discuss, in the conclusions section, 
how a similar result might be achieved by grid codes. 
While we believe this SI algorithm could be useful in real-space codes 
at least for certain types of problems, the reader interested in other
applications should proceed with caution.

Finally, we note that although our focus is in the plasma physics
field, the derivation of the iterative SI method presented is general and
should apply well to other stiff-wave physics phenomena that can be
described by means of advection-diffusion PDEs.

%
%
\section{Numerical scheme}
\label{num_test_sec}
\subsection{Derivation of the method}
\label{derivation_method}
As mentioned above, we are interested in dealing with coupled sets of
two-dimensional ad\-vect\-ion-diffusion partial differential equations. Let us consider the
following set as our general model:
\bea
\label{psi_eq_0}
\frac{\d \psi}{\d t}&=&\mathcal F(\phi,\psi)+
(\eta\nabla^2-\eta_H\nabla^4)\psi,\\
\label{phi_eq_0}
\frac{\d \phi}{\d t}&=&\mathcal G(\phi,\psi)+
(\nu\nabla^2-\nu_H\nabla^4)\phi.
\eea
Here, $\phi,\psi$ are arbitrary scalar fields, the operators $\FF, \GG$
represent the nonlinear advection-like operators, $\eta$ and $\eta_H$ are the resistivity and hyper-resistivity and $\nu$ and $\nu_H$ the viscosity and hyper-viscosity, respectively. In Fourier space, we can define the diffusion operators as:
\be
\mathcal D_{\eta}=\eta \kperp^2+\eta_H\kperp^4,\quad
\mathcal D_{\nu}=\nu \kperp^2+\nu_H\kperp^4
\ee
and thus rewrite the above set (in Fourier space), as:
\bea
\label{psi_eq}
\frac{\d \psi_k}{\d t}&=&\bar{\FF}(\phi,\psi)-\etak\psi_k,\\
\label{phi_eq}
\frac{\d \phi_k}{\d t}&=&\bar{\GG}(\phi,\psi)-\nuk\phi_k.
\eea
Here, the $k$ subscripts denote variables in Fourier space and $\bar{\FF},~\bar{\GG}$ represent the Fourier transforms of the nonlinear operators $\FF, \GG$. For notational simplicity, we will drop both the subscripts and the over bars: it is understood that we are working in Fourier space.
We note also that above we have chosen a fourth order operator to represent the hyperviscosity, but the following derivation holds irrespective of the order chosen for this operator.

The diffusion terms can be integrated analytically by means of the following variable transformations:
\be
\label{varsubs}
\psi=e^{-\etak t}\tilde\psi, \quad
\phi=e^{-\nuk t}\tilde\phi.
\ee
\Eqs{psi_eq}{phi_eq} thus become:	
\bea
\label{step1_psi}
\frac{\d\tilde\psi}{\d t}&=&e^{\etak t}\FF(\phi,\psi),\\
\label{step1_phi}
\frac{\d\tilde\phi}{\d t}&=&e^{\nuk t}\mathcal G(\phi,\psi).
\eea
We now apply a Crank-Nicolson discretization scheme (spatial indices
omitted for simplicity):
\bea
\label{step2_psi}
\frac{\tilde\psi^{n+1}-\tilde\psi^n}{\dt} &=&
   \frac{1}{2}\FF(\phi^n,\psi^n)
  +\frac{1}{2}e^{\etak \dt} \FF(\phi^{n+1},\psi^{n+1}),\\
\label{step2_phi}
 \frac{\tilde\phi^{n+1}-\tilde\phi^n}{\dt}&=&\frac{1}{2}\mathcal
   G(\phi^n,\psi^n) 
   +\frac{1}{2}e^{\nuk \dt}\mathcal G(\phi^{n+1},\psi^{n+1}).
\eea
(Without loss of generality, we have defined $t_n=0$, so $t_{n+1}=\dt$.)

Note that the handling of the diffusion terms through the variable
substitutions of (\ref{varsubs}) is a trivial step for spectral and
pseudo-spectral codes, but more complicated in real-space codes. However,
the derivation of the semi-implicit method that follows holds irrespective
of how the diffusion terms are handled -- this is only important for the
robust damping property, to be discussed later on in
\secref{robust_damp}. In \secref{conclusions} we briefly discuss how a similar treatment of the diffusive terms can be achieved in real space codes.

Usual implicit schemes would now invert \eqs{step2_psi}{step2_phi} to
determine the unknowns $\psi^{n+1}, \phi^{n+1}$ in terms of the known
quantities $\psi^{n},\phi^{n}$.  This may be done by, for example, an
iterative Jacobian-Free Newton Krylov solver.  A Newton method turns this
nonlinear root problem into a series of linear problems.  These linear
problems can often involve very large sparse matrices. If a direct
inversion was done, the very large timestep enhancements that can be
obtained relative to an explicit scheme might be severely offset by the
increased CPU time per step.  Jacobian-free Krylov iteration algorithms can
be used to solve the resulting linear problems without explicitly forming
the matrices.  If these iterative methods converge quickly enough (which
for hyperbolic problems may require very effective preconditioners), then
the implicit method can provide a significant improvement in CPU time
relative to an explicit algorithm.

The aim of our method, like more complicated iterative methods, is to
circumvent the need to directly invert a large linear problem.
We will use a simple iterative method, with a
semi-implicit operator providing an effective physics-based
preconditioner.  The first step is to reformulate
\eqs{step2_psi}{step2_phi} in an iterative way. To this effect, we begin by
rewriting these equations as:
\bea
\label{step3_psi}
\frac{\tilde\psi^{n+1,p+1}-\tilde\psi^n}{\dt}&=&\frac{1}{2} \FF(\phi^n,\psi^n)
+\frac{1}{2}e^{\etak \dt}\FF(\phi^{n+1,p+1},\psi^{n+1,p+1}),\\
\label{step3_phi}
\frac{\tilde\phi^{n+1,p+1}-\tilde\phi^n}{\dt}&=&\frac{1}{2}\mathcal G(\phi^n,\psi^n)
+\frac{1}{2}e^{\nuk \dt}\mathcal G(\phi^{n+1,p+1},\psi^{n+1,p+1}).
\eea
Next, we introduce one simple, but crucial, physics based assumption: that the stiff waves arise through the coupling of the equations and are well described by the assumptions 
\bea
\label{assumpts}
\FF(\psi^{n+1,p+1},\phi^{n+1,p+1})&\approx&\FF(\psi^{n+1,p},\phi^{n+1,p+1}),\nonumber\\
   &\approx& \FF(\psi^{n+1,p},\phi^{n+1,p}) + 
      \frac{\delta \FF}{\delta \phi} \( \phi^{n+1,p+1} - \phi^{n+1,p} \)
        ,\\
\label{assumpts2}
\GG(\psi^{n+1,p+1},\phi^{n+1,p+1})&\approx&\GG(\psi^{n+1,p+1},\phi^{n+1,p})
   \nonumber\\
   &\approx& \GG(\psi^{n+1,p},\phi^{n+1,p}) 
    + \frac{\delta \GG}{\delta \psi} \( \psi^{n+1,p+1} - \psi^{n+1,p} \).
\eea
In other words, we are assuming that the dependence of $\FF$ on $\psi$ is
relatively weak (i.e., the fast timescales arise via the dependence of $\FF$ on $\phi$, not on $\psi$) and can be approximated from the previous iteration $p$,
while it depends more sensitively on $\phi$ and so we need to keep its
dependence on the present $p+1$ iteration value by using a Taylor series 
(and \textit{vice-versa} for $\GG$).
This is certainly true in the linear limit for our model, as shown by
\eqs{linearized_ne}{linearized_psi}.  Most wave phenomena involve
oscillations between two different variables (for example, pressure and
velocity for sound waves, or position and velocity for a pendulum), so that
approximations like the above are often justified.  A key step when
extending this approach to more complicated systems of equations, which
might contain multiple waves, is to identify the key terms involved in the
fast wave dynamics.

[We have written the Jacobians in the above Taylor-series expansions as
functional derivatives, such as $\delta \FF / \delta \phi$, since $\FF$
may be a non-local functional that operates on $\phi$.  After spatial
discretization, $\delta \FF / \delta \phi$ will in general be a large
sparse matrix.  However, in our final algorithm all we have to evaluate
is an approximation to $(\delta \FF / \delta \phi)(\Delta \GG / \delta
\psi)$.  This approximation will be local in Fourier space, and so for
simplicity we will replace the functional derivative notation with the
standard notation $\partial \FF / \partial \phi$.  Note 
that in our final algorithm we never actually need to explicitly evaluate
the Jacobian, just the action of a Jacobian on a vector, and so this
formulation is ``Jacobian-free''.]  

With this assumption, the nonlinear future time term in \eq{step2_psi} can
be simplified by means of the Taylor expansion, i.e.:
\bea
\label{step4_psi}
\frac{\tilde\psi^{n+1,p+1}-\tilde\psi^n}{\dt}&=&\frac{1}{2} \FF(\phi^n,\psi^n)+\nonumber\\
& &\hspace{-2cm}+\frac{1}{2}e^{\etak \dt}\left[\FF(\phi^{n+1,p},\psi^{n+1,p})
+\frac{\d \FF}{\d \phi}(\phi^{n+1,p+1}-\phi^{n+1,p})\right].
\eea
An expression for the last term on the RHS of this equation is
straightforwardly obtained (for $p>0$) by taking the difference between 
\eq{step3_phi} evaluated for two successive values of $p$:
\be
\label{step4_phi}
\phi^{n+1,p+1}-\phi^{n+1,p}=\frac{\dt}{2}\frac{\d \GG}{\d \psi}\(\psi^{n+1,p+1}-\psi^{n+1,p}\),
\ee
where again we have used the approximations of \eq{assumpts}.
Undoing the variable substitution of \eq{varsubs}, \eq{step4_psi} thus becomes:
\bea
\label{step5_psi}
\psi^{n+1,p+1}=e^{-\etak \dt}\psi^n+& &\frac{\dt}{2}e^{-\etak \dt}\FF(\phi^n,\psi^n)
+\frac{\dt}{2}\FF(\phi^{n+1,p},\psi^{n+1,p})+\nonumber\\
& &+\frac{\dt^2}{4}\frac{\d \FF}{\d \phi}\frac{\d \mathcal G}{\d \psi}
(\psi^{n+1,p+1}-\psi^{n+1,p}).
\eea
At this point we note that if the above equation is iterated until
convergence, i.e., $|\psi^{n+1,p+1}-\psi^{n+1,p}|<\epsilon$, where
$\epsilon$ is the prescribed accuracy, the last term becomes
negligible. Therefore, it is not important to retain its exact functional
form as long as convergence is reached at each timestep. We can thus
simplify \eq{step5_psi} by replacing the last term on the RHS with an
analytically invertible approximation. We use:
\be
\label{step_approx}
\frac{\d \FF}{\d \phi}\frac{\d \mathcal G}{\d \psi}=-\hat\omega^2
\ee
where $\hat\omega$ is given by the dispersion relation, or an approximation
to it.  [Specific forms for the operator $\hat\omega^2$ will be discussed
later, for now it should just be thought of as an operator that is
relatively easy to invert.]

The steps we have just taken are similar to those for some previous SI
derivations and share some key advantages for high-frequency waves,
since it is only the combination $(\partial \FF / \partial \phi)(\partial
{\mathcal G}/\partial \psi)$ that has to be approximated by the
semi-implicit operator, and not $(\partial \FF / \partial \phi)$ and
$(\partial {\mathcal G}/\partial  \psi)$ separately.  As discussed in the
SI literature, this is much simpler because one does not have to worry
about getting the sign of the approximation $\hat{\omega}$ correct (i.e.,
one does not have to worry about the direction of propagation of the
waves). Chac\'on and Knoll~\cite{chacon_03,knoll_keyes_04,knoll_05} have
shown how this kind of physics-based semi-implicit operator can be related
to a Schur complement preconditioner, turning an originally hyperbolic
problem into a diagonally dominant parabolic problem that can be
efficiently solved by preconditioned Krylov (if the preconditioner is
sufficiently effective) and/or multigrid iterative techniques.  In some
cases, the resulting problem can also be efficiently solved with FFTs,
which we use here.

So far, we have: 
\bea
\label{step6_psi}
\psi^{n+1,p+1}&=& e^{-\etak \dt}\psi^n 
+\frac{\dt}{2}e^{-\etak \dt}\FF(\phi^n,\psi^n)
+\frac{\dt}{2}\FF(\phi^{n+1,p},\psi^{n+1,p})+\nonumber\\
& &-\frac{\hat\omega^2\dt^2}{4}(\psi^{n+1,p+1}-\psi^{n+1,p}),\\
\label{step6_phi}
\phi^{n+1,p+1}&=&e^{-\nuk\dt}\phi^n+\frac{\dt}{2}e^{-\nuk \dt}\GG(\phi^n,\psi^n)
+\frac{\dt}{2}\GG(\phi^{n+1,p},\psi^{n+1,p+1}).
\eea
\Eqs{step6_psi}{step6_phi} constitute the basis of our iterative
semi-implicit method. 
Since this iteration converges (as we will later show) to the Crank- Nicolson (CN) solution
for an arbitrary initial condition (within some domain), it is tempting to
just use these equations with the initial guess $\psi^{n+1,0}=\psi^n$,
$\phi^{n+1,0} = \phi^n$.  However, one finds that if one does this for an
undamped wave test problem, the amplitude of the solution converges to the
CN solution from above, so the algorithm is numerically unstable for any
finite number of iterations.  One can demonstrate this by considering the
test problem $\FF = \omega \phi$, $\GG= -\omega \psi$, $\etak=\nuk=0$,
finding that the squared amplification factor for the first iteration gives
$|A|^2 = ((\psi^{n+1,1})^2 + (\phi^{n+1,1})^2)/ ((\psi^{n})^2 +
(\phi^{n})^2) = 1 + 2 \omega^2 \dt^2 / (4 + \hat{\omega}^2 \dt^2)$.  For
high frequency modes that we are trying to treat implicitly, this becomes
$|A|^2 \approx 1 + 2 \omega^2 / \hat{\omega}^2$, which is significantly
larger than unity.  While further iterations would converge towards the
Crank-Nicolson solution, which has $|A|^2 = 1$, this significant numerical
amplification would still be problematic, and would require that the
semi-implicit operator be an extremely good approximation to the real
dynamics in order to converge quickly.   Furthermore, it is somewhat
bothersome that this algorithm would give such a large numerical
amplification, since it is supposed to be an approximation to the stable
Crank-Nicolson algorithm given by \eqs{step2_psi}{step2_phi}.
However, note \eqs{step6_psi}{step6_phi} were derived using \eq{step4_phi},
which is valid only for $p>0$, so an expression for the initial $p=0$ step
of the iteration requires further consideration.  It turns out that a more
careful treatment of the initial step will also make the final algorithm
stable.  Using \eq{assumpts2} in \eq{step6_phi} for $p=0$, we have:
\bea
\label{phi_1}
\phi^{n+1,1}=&&e^{-\nuk\dt}\phi^n+\frac{\dt}{2}e^{-\nuk \dt}\GG(\phi^n,\psi^n)+\nonumber\\
&&+\frac{\dt}{2}\[\GG(\phi^{n+1,0},\psi^{n+1,0})+\frac{\d\GG}{\d\psi}\(\psi^{n+1,1}-\psi^{n+1,0}\)\].
\eea
Let us define the quantity $\phi^{n+1,*}$ as:
\be
\label{phi_star}
\phi^{n+1,*}=e^{-\nuk\dt}\phi^n+\frac{\dt}{2}e^{-\nuk \dt}\GG(\phi^n,\psi^n)
+\frac{\dt}{2}\GG(\phi^{n+1,0},\psi^{n+1,0}),
\ee
which reduces \eq{phi_1} to:
\be
\label{phi_star_2}
\phi^{n+1,1}=\phi^{n+1,*}+\frac{\dt}{2}\frac{\d\GG}{\d\psi}\(\psi^{n+1,1}-\psi^{n+1,0}\).
\ee
Substituting this into \eq{step4_psi} evaluated at $p=0$, and using
\eq{assumpts2}, yields
%
%
\bea
\label{step8_psi}
\psi^{n+1,1}=& &e^{-\etak \dt}\psi^n
+\frac{\dt}{2}e^{-\etak\dt}\FF(\phi^n,\psi^n) \nonumber\\
&&+ \frac{\dt}{2}\FF(\phi^{n+1,*},\psi^{n+1,0})+
\frac{\dt^2}{4}\frac{\d \FF}{\d \phi}\frac{\d\GG}{\d\psi}\(\psi^{n+1,1}-\psi^{n+1,0}\),
\eea
where again the last term on the RHS can be simplified making use of
approximation (\ref{step_approx}).
Equation (\ref{step8_psi}) followed by \eq{step6_phi} (evaluated for $p=0$)
would give a 1st order accurate prediction of 
$(\psi^{n+1,1}, \phi^{n+1,1})$.  This could then be iterated once with
\Eqs{step6_psi}{step6_phi} to obtain a second order accurate calculation for
$(\psi^{n+1,2}, \phi^{n+1,2})$.  However, a small modification to the
prediction step can give a 2cd order accurate results for $(\psi^{n+1,1},
\phi^{n+1,1})$.
This is done by replacing $\psi^{n+1,0}$ in the nonlinear term of
\eq{step8_psi} with a 1st order accurate approximation, $\psi^{n+1,*}$, so
that \eq{step8_psi} becomes
\bea
\psi^{n+1,1}=& &e^{-\etak \dt}\psi^n
+\frac{\dt}{2}e^{-\etak\dt}\FF(\phi^n,\psi^n)+\nonumber\\
&&\frac{\dt}{2}\FF(\phi^{n+1,*},\psi^{n+1,*})-
\frac{\hat\omega^2\dt^2}{4}\(\psi^{n+1,1}-\psi^{n+1,0}\),
\eea
where $\psi^{n+1,*}$ is defined as:
\be
\label{psi_star}
\psi^{n+1,*}=e^{-\etak\dt}\psi^n+\frac{\dt}{2}e^{-\etak \dt}\FF(\phi^n,\psi^n)
+\frac{\dt}{2}\FF(\phi^{n+1,0},\psi^{n+1,0}).
\ee
Note that the replacement of $\FF(\phi^{n+1,*},\psi^{n+1,0})$ with
$\FF(\phi^{n+1,*},\psi^{n+1,1})$ is consistent with the assumption of
\eq{assumpts} that the dependence of $\FF$ on $\psi$ is relatively weak,
but is useful to ensure second order accuracy of the calculation of
$\psi^{n+1,1}$.   Similarly, when calculating $\phi^{n+1,0}$ from
\eq{step6_phi} for $p=0$, the term $\GG(\phi^{n+1,0},\psi^{n+1,1})$ is
replaced by $\GG(\phi^{n+1,*},\psi^{n+1,1})$.  This gives a 2cd order
accurate result for $\phi^{n+1,1}$.  The choice $\psi^{n+1,0}=\psi^n,
~\phi^{n+1,0}=\phi^n$ closes the specification of our semi-implicit scheme.

The final equations for the iterative SI algorithm can be written in a
compact form as follows:
\bea
\label{phi_star_final}
\phi^{n+1,*}&=&e^{-\nuk\dt}\phi^n+\frac{\dt}{2}\(1+e^{-\nuk \dt}\)\GG(\phi^n,\psi^n),\\
\label{psi_star_final}
\psi^{n+1,*}&=&e^{-\etak\dt}\psi^n+
\frac{\dt}{2}\(1+e^{-\etak \dt}\)\FF(\phi^n,\psi^n),\\
\label{psi_final}
\psi^{n+1,p+1}&=& e^{-\etak \dt}\psi^n +
\frac{\dt}{2}e^{-\etak \dt}\FF(\phi^n,\psi^n)
+\frac{\dt}{2}\FF(\phi^{n+1,p},\psi^{n+1,p})+\nonumber\\
& &-\frac{\hat\omega^2\dt^2}{4}(\psi^{n+1,p+1}-\bar\psi^{n+1,p}),\\
\label{phi_final}
\phi^{n+1,p+1}&=&e^{-\nuk\dt}\phi^n+\frac{\dt}{2}e^{-\nuk \dt}\GG(\phi^n,\psi^n)
+\frac{\dt}{2}\GG(\phi^{n+1,p},\psi^{n+1,p+1}).
\eea
where now $\phi^{n+1,0}=\phi^{n+1,*}$, $\psi^{n+1,0}=\psi^{n+1,*}$ and
$\bar\psi^{n+1,0}=\psi^n$, $\bar\psi^{n+1,p}=\psi^{n+1,p}$ for $p>0$.

We mentioned above that these equations ought to be iterated until
convergence, at which point the last term on the RHS of \eq{psi_final} is
negligible and these equations become equivalent to the fully implicit
\eqs{step2_psi}{step2_phi}. The relative importance of this term can be
quantified by: 
\be
\label{SI_error}
\mathcal E_{ji}^{p+1} = \left|\frac{\LL\(\psi_{ji}^{n+1,p+1}-\bar\psi_{ji}^{n+1,p}\)}
{\sqrt{\frac{1}{N}\sum_{ji}\left|\psi_{ji}^{n+1,p+1}-\psi_{ji}^n\right|^2}}\right|,
\ee
where $\LL = \hat{\omega}^2 \dt^2 /4$, $j,i$ are the Fourier (or real) space
grid point indexes and $N$ is the total number of grid points.  The
timestep of the integration and/or the number of $p$ iterations is thus
determined by the requirement:
\be
\label{SI_error_cond}
\max|\mathcal E_{ji}^{p+1}|<\epsilon
\ee
where $\epsilon$ is the user prescribed accuracy.  
We emphasize that this error is a measure of the convergence of the iteration scheme to the Crank-Nicolson difference equation. It is not a direct 
control of the overall error in the integration (which,
when converged, will be that of a CN scheme).
For convenience, we also
define the total number of iterations of \eqs{psi_final}{phi_final} as
$\pmax$.  I.e., $\pmax$ is the number of times the semi-implicit operator
is inverted in corrector steps.  $\pmax=1$ evaluates
\eqs{psi_final}{phi_final} once, for $p=0$, to determine $(\psi^{n+1,1},
\phi^{n+1,1})$.  $\pmax=2$ evaluates \eqs{psi_final}{phi_final} twice, for
$p=0, 1$, and calculates up to $(\psi^{n+1,2}, \phi^{n+1,2})$.

There are several variations of SI methods and their derivations in the
literature, some using a more heuristic approach of adding and subtracting
an operator to the RHS of one of the equations, and some taking a more
systematic approach to deriving an SI operator.  Also, some SI algorithms
have used a leap-frog-like staggered time grid.  Our final result is
closest to certain SI methods using non-staggered grids.  For example, our
\eqs{phi_star_final}{phi_final} for $\pmax=1$ are equivalent to Eqs.(11-14)
of \cite{harned_86}.  The formulation presented here makes it clear how to
extend other semi-implicit methods to use additional iterations if desired
(in this regard, note that it is important that the definition of
$\bar{\psi}^{n+1,p}$ is different on the first $p=0$ corrector step than in
later iterations).  These additional iterations provide a simple way to
turn a semi-implicit method into a fully implicit method, and further
demonstrate the relation between SI ideas and preconditioning for
iterative methods, as discussed in \secref{equiv_precond}.
\subsection {Linear stability analysis}
\label{lin_stab_analysis}
We will now show under which conditions the method described by \eqs{phi_star_final}{phi_final} is unconditionally stable with respect to the timestep $\dt$. For simplicity, we neglect the diffusive terms (since these terms are treated implicitly, they have a stabilizing effect). 
In line with the assumptions of (\ref{assumpts}), we assume that, in the linear regime, the following is true:
\bea
\label{linear_ops}
\FF(\phi,\psi)&\approx& f\phi,\nonumber\\
\GG(\phi,\psi)&\approx& g\psi.
\eea
Taking the simplest case of $p_{max}=1$ (i.e., a predictor-corrector
scheme) \eqs{phi_star_final}{phi_final} can be written in matrix form as:
\be
\[
\begin{array}{cc}
1 & -\dt/2~g\\
0 & 1
\end{array}
\]
\[
\begin{array} {cc}
\phi^{n+1,1}\\
\psi^{n+1,1}
\end{array}\]
=\[
\begin{array}{cc}
1 & \dt/2~g\\
\frac{\dt f }{1+\LL} & 
1+\frac{\dt^2 f g}{2(1+\LL)}
\end{array}
\]
\[
\begin{array} {cc}
\phi^{n}\\
\psi^{n}
\end{array}\],
\ee
where $\LL = \hat{\omega}^2 (\dt)^2/4$.  Inversion of the matrix on the LHS
yields the equation 
\be
\[
\begin{array} {cc}
\phi^{n+1,1}\\
\psi^{n+1,1}
\end{array}\]
=
A
\[
\begin{array} {cc}
\phi^{n}\\
\psi^{n}
\end{array}\],
\ee
where $A$ is the amplification matrix:
\be
A=
\[
\begin{array}{cc}
 1+\frac{f g \dt^2}{2 \LL+2} & \frac{\dt g \left(f g \dt^2+4 \LL+4\right)}{4 (\LL+1)}
   \\
 \frac{\dt f}{\LL+1} & 1+\frac{f g \dt^2}{2 \LL+2}
\end{array}
\].
\label{A_matrix}
\ee
The method is unconditionally stable when the eigenvalues $\lambda_{\pm}$ of the matrix $A$ are such that $|\lambda_\pm|\leq1$. Using $fg=-\omega^2$, we obtain:
\be
\label{eigenvalues}
\lambda_{\pm}=\frac{4+\left(\hat\omega^2-2 \omega ^2\right) \dt^2\pm2 i \omega \dt	 \sqrt{4+\left(\hat\omega^2-\omega ^2\right)
   \dt^2}}{4+\dt^2 \hat\omega^2}
\ee
It is straightforward to show that this algorithm is symplectic,
$|\lambda_\pm| = 1$, if the argument of the square root is positive.  Thus
we have unconditional stability for arbitrarily large $\omega \dt$ if
$\hat{\omega}^2 \dt^2 > \omega^2 \dt^2 -4$, or in the relevant limit of
$\omega \dt \gg 1$, if $\hat{\omega}^2 > \omega^2$.  
Since the magnitude of the error at
each iteration, expression (\ref{SI_error}) is proportional to
$\hat\omega^2$, it is clear that setting $\hat\omega^2 \approx\omega^2$ (but
$\hat\omega^2>\omega^2$) is the best choice for stability and accuracy.
(Of course, nonlinearly $\omega^2$ is not known, so one wishes to
set $\hat{\omega}^2$ to be an approximate upper bound on the true
operator $\omega^2$.)   Here we have demonstrated that the first
corrector step, $\pmax=1$, gives a solution that is symplectic (at least
for this linear wave test problem).  Additional iterations would eventually
converge to the Crank-Nicolson solution, which is also symplectic.  We have
also proven that the algorithm is symplectic for arbitrary $\pmax$, as
discussed at the end of the next section.

Finally, note that a Taylor series expansion of expressions
(\ref{eigenvalues}) yields:
\be
\lambda_{\pm}=1\pm i \omega \dt -\frac{\omega^2\dt^2}{2}\pm \mathcal O (\dt^3),
\ee
consistent with the fact that the first order accurate predictor step of
\eqs{phi_star_final}{psi_star_final} ensures that the corrector iterations
are 2cd order accurate.

\subsection{Linear convergence rate}
\label{lin_conv_rate}
To determine the convergence rate of the iterative scheme presented in \eqs{phi_star_final}{phi_final} we set:
\bea
\phi^{n+1,p+1}&=&\phi^{n+1}+\delta \phi^{n+1,p+1},\\
\psi^{n+1,p+1}&=&\psi^{n+1}+\delta \psi^{n+1,p+1}.
\eea
where the first term on the RHS is the converged solution and $\delta \phi^{n+1,p+1},\delta \psi^{n+1,p+1}$ are the error at each iteration. Again using the linear approximations of equations (\ref{linear_ops}), it is straightforward to obtain the equation
\be
\[
\begin{array} {cc}
\delta \phi^{n+1,p+1}\\
\delta \psi^{n+1,p+1}
\end{array}\]
=\[
\begin{array}{cc}
 \frac{\dt^2 f g}{4 (\LL+1)} & \frac{\dt g \LL}{2 (\LL+1)} \\
 \frac{\dt f}{2 (\LL+1)} & \frac{\LL}{\LL+1}
\end{array}
\]
\[
\begin{array} {cc}
\delta \phi^{n+1,p}\\
\delta\psi^{n+1,p}
\end{array}\].
\label{convergence_matrix}
\ee
The matrix on the RHS has the eigenvalues
\be
\lambda_1=0,\quad 
\lambda_2=\frac{\dt^2\(\hat\omega^2-\omega^2\)}{4+\hat\omega^2\dt^2}.
\label{convergence_eigenvalues}
\ee
Clearly $\lambda_2$ is the eigenvalue of interest. In the limit of $\hat\omega \dt \gg 1$, this expression simplifies to:
\be
\label{eigen_conv}
\lambda_2=1-\frac{\omega^2}{\hat\omega^2}.
\ee

We showed in the previous section that $\hat\omega^2\approx\omega^2$ is the
best choice in terms of achieving stability while minimizing the error
(\ref{SI_error}) at each iteration. Expression (\ref{eigen_conv}) shows
that this is also the best choice in order to maximize the convergence
rate. 
%
%
However, the iteration will converge as long as the semi-implicit operator
is large enough.  The stability condition for the first iteration,
$\hat\omega^2 \dt^2 > \omega^2 \dt^2 - 4$ from the previous section, is also
a sufficient condition for convergence.

In the previous section, we demonstrated that the algorithm for $\pmax=1$
is symplectic.  We have also shown this is true for arbitrary $\pmax$, and
here we summarize the main steps.  First write \Eq{convergence_matrix},
which holds for $\pmax \ge 1$, in the vector-matrix form $\delta \vec
y^{n+1,p+1} = \mat B \, \delta \vec y^{n+1,p} = (\mat{B})^p \, \delta \vec
y^{n+1,1}$, where $(\mat{B})^p$ denotes the matrix $\mat{B}$ raised to
the $p$'th power (as opposed to an index superscript).  Expressing the
converged Crank-Nicolson solution as $\vec{y}^{n+1} = \mat C \vec y^n$
(where the matrix $\mat C$ can be easily calculated), and the $\pmax=1$
solution as $\vec y^{n+1,1} = \mat A \vec y^n$ (where $\mat A$ is given by
\Eq{A_matrix}), this can be written as $\vec y^{n+1,p+1} = \left[ \mat C +
(\mat B)^p (\mat A - \mat C) \right] \vec y^n \doteq \mat A_p \vec y^n$.
The eigenvalues of the matrix $\mat A_p$ can then be calculated using a
symbolic mathematics package for simplicity (we used Maple~\cite{Maple} for
this calculation). To reduce the complexity of intermediate equations,
it is useful to express $(\mat B)^p = \mat E (\mat D)^p \mat E^{-1}$, where
the columns of $\mat E$ are the eigenvectors of $\mat B$ and the diagonal
matrix $(\mat D)^p = (0, \, 0 ;\, 0, \, (\lambda_2)^p)$ (without yet
substituting the actual value of the eigenvalue $\lambda_2$ from
\Eq{convergence_eigenvalues}).  One can then show that the eigenvalues of
$\mat A_p$ can be written in the form $\lambda_{p,\pm} = c_1 \pm i
\sqrt{c_2}$, where $c_1$ is real and a sufficient condition that $c_2$
be positive is again the stability criterion from the previous section,
$\hat\omega^2 \dt^2 > \omega^2 \dt^2 - 4$.  The terms $c_1$ and $c_2$
satisfy $c_1^2 + c_2=1$, thus showing that this algorithm is symplectic
for an arbitrary number of iterations $\pmax$.

\subsection{On the relation between the iterative semi-implicit method
and a preconditioned Jacobian-Free Newton iterative method}
\label{equiv_precond}
To investigate some of the properties of this algorithm, let us for
simplicity consider now a more compact vector form provided by the
following equation:
\be
\frac{\partial \psi}{\partial t} = {\mathcal F}(\psi) - \DD_\eta \psi,
\label{compact_eq}
\ee
where $\psi=\psi_i(\vec x, t)$ may be a vector field (with components
$i=1..N$), ${\mathcal F}$ is a nonlinear operator, and $\DD_\eta$ is a linear
operator (which represents dissipation in our problems).  Introducing the
variable  transformation $\psi(t) = e^{-\DD_\eta t} \tilde{\psi}(t)$, this
becomes 
\be
\frac{\partial \tilde \psi}{\partial t} = e^{\DD_\eta t} {\mathcal F}(\psi)
\doteq \tilde{\mathcal F}(\tilde \psi).
\label{compact_eq2}
\ee
(A definition is denoted by $\doteq$.)  For simplicity we will drop the
tildes over variables in the rest of this subsection.
It is not hard to see that, in this case, the numerical scheme of
\eqs{psi_final}{phi_final} (using \eq{assumpts} to approximate $\FF$) 
is equivalent, for $p>1$, to:
\bea
\label{psi_simple}
\psi^{n+1,p+1}&=& \psi^n +
\frac{\dt}{2}\[\FF(\psi^n)+\FF(\psi^{n+1,p})+
\hat \FF(\psi^{n+1,p+1}-\psi^{n+1,p})\],
\eea
where $\hat \FF$ is the semi-implicit operator, and therefore, chosen to be
invertible.   
\Eq{psi_simple} can thus be easily manipulated to yield:
\be
\label{psi_simple_inv}
\psi^{n+1,p+1}=\(1-\frac{\dt}{2}\hat\FF\)^{-1}\[\psi^n+\frac{\dt}{2}\FF(\psi^n)
+\frac{\dt}{2}\FF(\psi^{n+1,p})
-\frac{\dt}{2}\hat\FF(\psi^{n+1,p}) \]
\ee
We now show how the above equation can be obtained with a Jacobian-free
Newton algorithm coupled with a preconditioned simple functional iteration.
A Crank-Nicolson discretization of \eq{compact_eq2} leads to:
\be
\psi^{n+1}-\frac{\dt}{2}\FF(\psi^{n+1})=\psi^n+\frac{\dt}{2}\FF(\psi^n)
\ee
As mentioned before, directly inverting the LHS of this equation is often
hard and impractical. A common and useful approach is to resort to
preconditioned iterative methods.  For nonlinear cases like this, it is
useful to use an outer Newton iteration for the nonlinear part of
the problem, and an inner iteration with preconditioning for the resulting
linear part of the problem.  I.e., introducing a Newton iteration count
$p$, the above equation becomes:
\be
\psi^{n+1,p+1}-\frac{\dt}{2}\left[ \FF(\psi^{n+1,p})
+ \frac{\delta\FF}{\delta \psi} (\psi^{n+1,p+1} - \psi^{n+1,p}) \right]
=\psi^n+\frac{\dt}{2}\FF(\psi^n),
\label{compact_Newton}
\ee
Because $\delta \FF / \delta \psi$ is a complicated, large sparse matrix,
it is still impractical to directly solve this problem.   If we have an
approximation to $\delta \FF / \delta \psi$, denoted by $\hat \FF$, that is
not too hard to invert, we can use this as a preconditioner, and write
\eq{compact_Newton} as
\bea
\left(1 - \frac{\Delta t}{2} \hat\FF \right)^{-1} & &
\left(1 - \frac{\Delta t}{2} \frac{\delta \FF}{\delta \psi} \right)
\psi^{n+1,p+1} = \nonumber \\
\left(1 - \frac{\Delta t}{2} \hat\FF \right)^{-1} & & 
\left[ \psi^n +\frac{\dt}{2}\FF(\psi^n) + \frac{\dt}{2}\FF(\psi^{n+1,p})
- \frac{\Delta t}{2} \frac{\delta \FF}{\delta \psi} \psi^{n+1,p} \right].
\eea

This is of the form $\hat{A}^{-1} A \psi^{n+1,p+1} = b^p$.  If the
preconditioner is sufficiently good, then $\hat{A}^{-1} A$ is close enough
to the identity matrix that a simple functional iteration should be 
sufficient.  Using an index $q$ to label these iterations within
each Newton step, this would become $\psi^{n+1,p+1,q+1} = b^p +
(1-\hat{A}^{-1} A) \psi^{n+1,p+1,q}$, with the initial condition
$\psi^{n+1,p+1,0}=\psi^{n+1,p}$.  This should converge as long as the 
magnitude of the eigenvalues of $(1-\hat{A}^{-1} A)= \hat{A}^{-1}(\hat{A} -
A)$ are all less than unity.  If we do only one inner functional iteration
($q=0$) per outer Newton iteration, then this exactly reproduces
\eq{psi_simple_inv}.   [Alternatively, replacing $\delta \FF / \delta \psi$
in \eq{compact_Newton} with the approximate Jacobian $\hat \FF$ leads to
\eq{psi_simple_inv}.]  The Iterative Semi-Implicit algorithm presented in
this paper is essentially this, with a particular physics-based form for
$\hat \FF$ appropriate for high frequency waves arising from coupling
between differential equations, and a modification of the first step to
give a symplectic second-order accurate result after one corrector step.

More advanced iterative methods could be used, such as Krylov methods like
GMRES or restarted Loose GMRES~\cite{baker05}, and will accelerate
convergence for some types of problems.
(Also, because of the various approximations made, our simplified algorithm
will not have the asymptotic quadratic convergence of a full Newton
algorithm on the nonlinear part of the problem.)
However, we have found that simple
functional iteration works quite well for our problems, converging in just
an iteration or two.  In part, this is because there is very little energy
contained in the high frequency waves in these problems.  In principle,
convergence might be slow if there were a significant component of the
error vector in these high frequency waves, for which
$\hat{A}^{-1}(\hat{A}-A)$ has the largest eigenvalues.  But there is very
little physical coupling between the low frequency dynamics of interest and
the high frequency waves, so if they are treated in a numerically stable
way, they will have been damped away on previous time steps and do not
require many iterations on subsequent time steps.

Although we use a simple functional iteration, note that the optimal method
for inverting the physics-based preconditioner $(1 -\frac{\dt}{2}\hat\FF)$
itself might in some cases be an iterative algorithm (more on this in
\secref{semi-implicit_operator}), or in other cases with FFTs or a direct
sparse matrix solver.  For our application, we are already using FFTs for a
pseudo-spectral evaluation of the nonlinear terms, so it is easy to
implement and very efficient to directly invert our physics-based
preconditioner in wave-number space.

\subsection{Second order operator splitting}
\label{sec_ord_op_split}
In the derivation of the numerical scheme of \secref{derivation_method},
the nonlinear advection operators and the diffusion operators are treated
differently. Operator splitting methods can give rise to splitting errors
in numerical schemes, resulting in an overall order of the scheme which is
lower than that of any of the individual methods applied to treat each of
the operators. In this section, we show that our scheme is second order
accurate even if these operators do not commute. To demonstrate this point,
it is sufficient to consider the paradigm posed by the compact vector form
of \eq{compact_eq}.  Crank-Nicolson discretization of \eq{compact_eq2}
yields
\be
\tilde{\psi}^{n+1}-\tilde{\psi}^n =
  \frac{\dt}{2} e^{\DD_\eta(t+\dt)} \FF(\psi^{n+1}) 
+ \frac{\dt}{2} e^{\DD_\eta t}      \FF(\psi^{n}).
\ee
Undoing the variable transformation and rearranging gives
\be
\psi^{n+1} - \frac{\dt}{2} \FF(\psi^{n+1}) = e^{-\DD_\eta \dt} 
   \(\psi^{n} + \frac{\dt}{2} \FF(\psi^n) \).
\label{sec_order}
\ee
Through second-order accuracy, the LHS can be expanded as
LHS=$\psi^{n+1}-(\dt/2)\[\FF(\psi^n) + (\partial \FF / \partial
\psi)(\psi^{n+1} \right. \linebreak[1] \left.-\psi^n)\]$.  Then the above
equation becomes
\be
\(1 - \frac{\dt}{2} \frac{\partial \FF}{\partial \psi} \) \psi^{n+1} 
= e^{-\DD_\eta \dt} \(\psi^{n} + \frac{\dt}{2} \FF(\psi^n) \) 
+ \frac{\dt}{2} \FF(\psi^n) - \frac{\dt}{2} \frac{\partial \FF}{\partial
  \psi} \psi^n
\label{splitting_expansion}
\ee
From a second order Taylor-series expansion, $\psi(t+\dt) = \psi(t) + \dt
\psi' + \frac{1}{2} \dt^2 \psi''$, we see that the solution to
\eq{compact_eq} through second order accuracy should be:
\bea
\psi^{n+1} &=& \psi^n + \dt \FF(\psi^n) - \dt \DD_\eta \psi^n \nonumber \\
& &  + \frac{\dt^2}{2} \[ \frac{\partial \FF}{\partial \psi} \FF(\psi^n) 
    - \frac{\partial \FF}{\partial \psi} \DD_\eta \psi^n 
    - \DD_\eta F(\psi^n) + \DD_\eta^2 \psi^n \].
\label{Taylor_series2}
\eea 
Inverting the operator on the LHS of \eq{splitting_expansion} and
expanding, one can show that \eq{splitting_expansion} agrees with
\eq{Taylor_series2} through second order.  Thus we have demonstrated that,
at least in the limit of $\dt\d\FF/\d \psi\ll1$,
our splitting scheme is second order accurate even if the advection ($\FF$)
and the diffusion ($\DD_\eta$) operators do not commute.
Numerical demonstration of the second order convergence of the algorithm
for our test problem will be 
presented in \secref{num_tests_sec}, where it will be seen that the 
algorithm remains second order accurate even at values of $\dt$ considerably 
larger than those required by an explicit integration.
\subsection{Robust damping}
\label{robust_damp}
\begin{figure}
\begin{center}
\epsfig{file=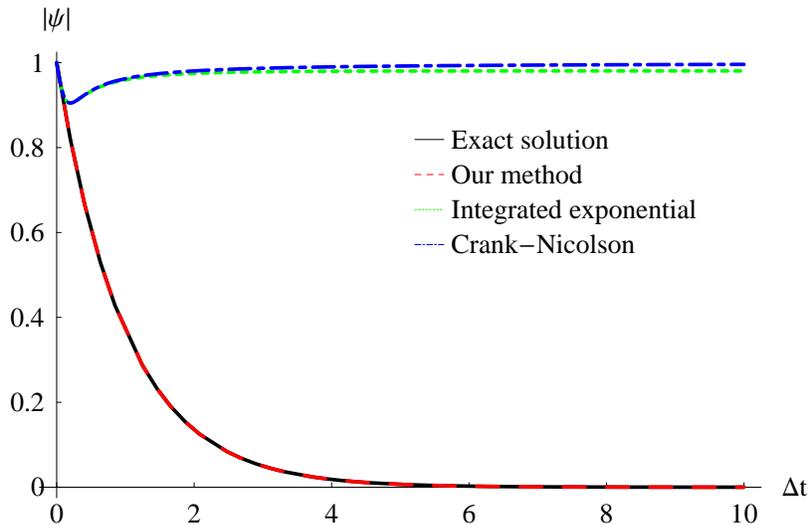,width=12cm} 
\end{center}
\caption{\footnotesize{Plot of the solution of \eq{robust_eq} as obtained
analytically, expression (\ref{robust_exact}), and with three different
numerical methods. Input parameters are $\omega=10, \DD_\eta=1$ and $\psi_0=1$.}}
\vspace{0.5cm}
\label{robust_pic}
\end{figure}

Let us consider a simple test problem based on a linear scalar version of
\eq{compact_eq} with $\FF=-i \omega$ and $\DD_\eta$ a real number,
\be
\label{robust_eq}
\frac{d \psi}{d t}=-(i\omega+\etak )\psi.
\ee
The analytic solution is straightforward:
\be
\label{robust_exact}
\psi=\psi_0e^{-(i\omega+\etak)t},
\ee
where $\psi_0$ is the initial condition. A plot of the magnitude of this
solution is shown in~\fig{robust_pic}, for the parameters
$\omega=10,~\psi_0=1$ and $\etak=1$ (full black line). Also shown are the
numerical solutions of \eq{robust_eq} with three different numerical
methods: in red (dashed line), the solution yielded by \eq{sec_order}
(i.e., the method described in this paper) 
$\psi^{n+1} = (1+i \omega \dt / 2)^{-1}e^{-\DD_\eta \dt} (1 - i \omega \dt
/2) \psi^n$ ; in green (dotted line) the result obtained by the analytic
integration of the diffusive exponential, as described in
appendix~\ref{exact_exp} and, in blue (dash-dotted line), the solution
given by a Crank-Nicolson discretization of the RHS. As seen, for
$\omega \dt<1$, all three methods reproduce the exact solution. However,
only the discretization of \eq{sec_order} is capable of capturing the
behavior of the exact solution for \textit{any} value of $\dt$, as is
desirable for a robust treatment of damping with implicit integration,
where $\etak\dt\gg1$.  One might try to improve the CN approach by using CN
for the wave term and backwards Euler for the damping term, i.e.,
discretize \eq{robust_eq} as $(\psi^{n+1}-\psi^n)/\dt = - i \omega
(\psi^{n+1} + \psi^n) /2 - \etak \psi^{n+1}$.  However, in the limit $\dt
\rightarrow \infty$ this gives $\psi^{n+1}/\psi^n = - \omega / ( \omega - i
2 \DD_\eta)$, which for $\DD_\eta \ll \omega$ gives very little damping, even
though $\DD_\eta \dt >> 1$.

The algorithm presented here preserves the property of robust damping,
which mathematically is termed $L$-stability.  An algorithm that is stable
for the \eq{robust_eq} for arbitrary $\Delta t$ (for $\DD_\eta \ge 0$) is
called $A$-stable.  An algorithm is $L$-stable if it is $A$-stable and has
the property of the exact solution that $\psi^{n+1}/\psi^n \rightarrow 0$
as $\dt \rightarrow \infty$.  It is well known that Crank-Nicolson is
$A$-stable but not $L$-stable.  In physical terms, while Crank-Nicolson has
the nice property of giving no artificial damping of pure wave problems
(i.e., it is symplectic for $\DD_\eta = 0$), it has a difficulty in not
giving significant damping for problems that should be strongly damped in
the large $\dt$ limit.  Examples of $L$-stable algorithms include the 2cd
order Backward Differentiation Formula (BDF2), and certain Rosenbrock or
Implicit-Explicit Runge-Kutta algorithms.  The iterated semi-implicit
method presented here is based on a kind of integrating-factor variation of
Crank-Nicolson that makes it $L$-stable. (An algorithm based on the Modified
Crank-Nicolson algorithm~\cite{Ascher95,Ruuth95} could be another way to
improve the treatment of damping.)  Alternatively, one could develop an
iterative semi-implicit algorithm starting from BDF2 instead of CN.  This
would be automatically $L$-stable and would treat the damping and other
terms in a balanced way that might have certain advantages for some
applications.  On the other 
hand, BDF2 introduces a weak numerical damping of resolved waves,
$|\psi^{n+1}/\psi^n| \approx (1 - (\omega \dt)^4/4 + {\mathcal O}(\dt^6))$,
while the algorithm presented here has the advantage of retaining the CN
property of being dissipation free in the limit $\DD_\eta = 0$.
Future work could investigate the relative advantages of these different
approaches.

%
\section{Application -- gyrofluid magnetic reconnection}
\label{model}
The field of plasma physics is particularly rich in stiff wave
problems. One such application is magnetic
reconnection~\cite{bisk_MR,priest_forbes}, a phenomenon in which magnetic
field lines that are being convected with the plasma flows suddenly break
and reconnect in a different configuration. Magnetic energy is released in
this process, giving rise to high velocity plasma flows, energetic
particles and plasma heating. Reconnection is the cause of solar
flares~\cite{sweet_review, yokoyama_01}; it is also manifest in the
interaction between the solar wind and the Earth's magnetic
field~\cite{dungey_61} in the magnetopause~\cite{frey_03} and
magnetotail~\cite{xiao_06}. In fusion devices it plays a crucial role in
the development of the sawtooth instability, which can be detrimental to
the plasma confinement~\cite{hastie_97}.

In the absence of external forcing, magnetic reconnection arises as a
result of a well known plasma instability called the tearing mode
\cite{FKR}. 
Regardless of the physical framework [i.e., resistive magnetohydrodynamics (MHD), Hall-MHD\footnote{Hall-MHD consists in extending the usual MHD Ohm's law to include the so-called Hall term, $\bm J \times \bm B$.}, kinetic, etc.], this instability is particularly hard to simulate numerically (see, e.g., ~\cite{tang_05, chacon_03}). The reason is 
essentially one: the intrinsically multiscale character, both temporal and spatial, of magnetic reconnection. With respect to its time evolution, the problem arises because the typical growth rate $\gamma$ of the tearing instability is much smaller than the frequency of the fastest waves present in the system (e.g., Alfv\'en waves, Kinetic Alfv\'en waves, whistler, etc.); spatially, this problem requires the simulation of at least two, but often more, spatially distinct scales: the large scale equilibrium and the short scale dissipation layer. The separation between these two scales is a function of the magnitude of the field-line breaking terms (e.g., resistivity, electron inertia) which, in order to address a physically meaningful region of parameter space, have to have extremely small magnitudes, and thus become important only in very narrow regions in the plasma where strong gradients arise. 
It follows from these conditions that numerical simulations of magnetic reconnection require extremely high spatial resolutions and either an implicit time integration scheme or, if an explicit scheme is adopted, timesteps which are much smaller than $1/\gamma$.

In many plasmas of interest, such as those in fusion devices, or at the magnetotail or solar flares, it has long been suspected that an MHD description is too simple to fully explain the complexity of the observations. The discrepancy is obvious in the observed and calculated reconnection rates which, in the MHD framework, differ by several orders of magnitude. Interest has diverged to non-MHD effects which might cause a speed-up of this process~\cite{aydemir_92,GEM_birn,malyshkin_05}. The potential candidates differ depending on the specific geometry and plasma parameters.
For example, in most present fusion plasmas, the ion Larmor radius exceeds, or at least is comparable to, the width of the dissipation region. 
Thus, finite ion Larmor radius (FLR) effects cannot be neglected and are, in fact, known to produce a speed up of the reconnection rate \cite{aydemir_92,kleva_95}. 
From the numerical point of view, the FLR terms bring about one further
complication: their inclusion in the equations fundamentally changes the
dispersion relation of the shear Alfv\'en wave, essentially generalizing it
to the so-called Kinetic Alfv\'en wave (KAW), which has a high frequency dispersive character, i.e.,  $\omega\sim\kperp^2$, where $\omega$ is the wave frequency and $\kperp$ the wavenumber (the counterpart to the FLR effects in the absence of a strong magnetic field is the Hall term in Ohm's law, which also introduces a high frequency dispersive wave, the whistler -- see \cite{rogers_01} for a detailed discussion of the relative roles of these waves).
Explicit numerical integration schemes are forced to resolve the highest
frequencies present in the system, i.e., they must integrate on a
timestep $\dt\sim1/\omega_{max}$. Therefore, when dispersive waves are
present, $\dt\sim1/k_{\perp, max}^2$. Since $k_{\perp, max}\sim 1/\Delta
x $, where $\Delta x$ is the grid-spacing, the time step becomes
impractically small as higher resolutions are required to resolve the
extremely thin reconnection layer.
Reconnection studies with FLR or Hall terms typically use
hyper-diffusion (or hyper-resistivity) to limit how thin the
reconnection layer becomes, which helps limit how small the grid spacing
$\Delta x$ needs to be, and thus allows the time step $\Delta t$ to be
larger than it would be otherwise.  There can be a certain amount of physical
hyper-resistivity in the Ohm's law due to electron viscosity, and the
macroscopic physical results are found to be independent of the choice
of hyper-diffusion over a range of small hyper-diffusion, but if the
hyper-diffusion is too large it can affect some of the physical
results of interest.  For typical values of hyper-diffusion used in
simulations, the required grid resolution is still fairly fine and the
resulting explicit time step limitation is still quite severe, so one is
motivated to search for an implicit or semi-implicit algorithm to bypass
this severe time step limitation.

%

\subsection{Equations}
To study FLR effects in magnetic reconnection we chose a 2 field
gyrofluid model, described by the following normalized equations:
\bea
\label{gyro_ne_1}
\frac{\d n_e}{\d t}+[\phi , n_{e}]&=&[\psi,\nabperp^2 \psi]+
\nu\nabperp^2 n_e-\nu_H\nabperp^4n_e,\\
\label{gyro_psi_1}
\frac{\d \psi}{\d t}+[\phi , \psi]&=&\rho_{s}^{2}[n_e, \psi]+
\eta \nabperp^{2} (\psi-\psi_{eq})-
\eta_H \nabperp^{4} (\psi-\psi_{eq}),\\
\label{gyro_poisson}
n_e&=&\frac{1}{\roi^{2}}\[\hat\Gamma _{0}(b)-1\]\phi.
\eea
Here $n_e$ is the perturbed electron density, $\phi$ represents the electrostatic potential, $\psi$ is the magnetic flux, related to the in-plane magnetic field by $\Bperp=\ez \times \bm \nabla \psi$,  $[P,Q]=\d_{x}P\d_{y}Q-\d_{y}P\d_{x}Q$ is the Poisson bracket and $\roi$ and $\ros$ are the ion and the ion-sound Larmor radius, respectively. Auxiliary quantities are the parallel current density, $j_\parallel=\nabperp^2\psi$, where $\nabperp^2=\d_x^2+\d_y^2$,
and the flow velocity, $\bm v_\perp=\ez \times \bm \nabla \phi$. Dissipation is provided by the viscosity $\nu$ and the resistivity $\eta$. We also add, for physical reasons and numerical convenience, phenomenological hyper-diffusion coefficients, $\nu_H$ and $\eta_H$~\cite{rogers_96,breslau_03}. In \eq{gyro_psi_1}, the equilibrium is subtracted from the nonideal terms on the RHS in order to impose an ideal equilibrium [this is equivalent to imposing an external electric field, $E_{ext}=\(\eta\nabperp^2-\eta_H\nabperp^4\)\psi_{eq}$] -- see appendix~\ref{eq_term} for how to include this term in the numerical scheme of \eqs{phi_star_final}{phi_final}. Time is normalized to the Alfv\'en time, lengths to the equilibrium scale length.

These equations are reminiscent of the more complete models of 
Snyder and Hammett~\cite{snyder_01}
and Schep \etal~\cite{schep_94}. 
\Eq{gyro_poisson} is known in the literature as the gyrokinetic Poisson equation \cite{lee_83}. The integral operator $\hat\Gamma_0$ expresses the average of the electrostatic potential over rings of radius $\roi$. In Fourier space, this operator simply becomes
\be
\Gamma _{0}(b)=e^{-b}I_{0}(b),
\ee
where $b=k_{\perp}^{2} \rho _{i}^{2}$ and  $I_{0}$ is the modified Bessel function of zeroth order. In other words, within the gyrokinetic work frame, the particle is allowed to experience different values of the electrostatic potential as it orbits the magnetic guide field. The purpose of the $\Gamma_0$ operator is to average $\phi$ over such orbits and over the Maxwellian velocity distribution. When implemented numerically in real space, a Pad\'e approximant of this operator is often used~\cite{hammett_92}:
\be
\Gamma_0(b)-1\approx-\frac{b}{1+b},
\ee
which converts \eq{gyro_poisson} into:
\be
\label{pade_gyro}
\(1-\roi^2\nabperp^2\)n_e=\nabperp^2\phi.
\ee
In our case, \eq{gyro_poisson} is solved in Fourier space and thus no
approximation to $\Gamma_0(b)$ is required. In the limit $k_\perp
\rho_i\rightarrow0$, the above set of equations reduces simply to the
Reduced MHD (RMHD) model of Strauss~\cite{strauss_76}.

Sets of fluid equations similar to the above, or their collisionless
version, which further include electron inertia, have been in use for quite
some time to study magnetic reconnection (e.g. Kleva
\etal~\cite{kleva_95}). The particular form presented, however, differs
from most in that it also includes the effect of finite ion temperature
(i.e., $\roi\neq0$) and is unique, so far, in magnetic reconnection
studies, in keeping the full form of the Gyrokinetic Poisson law,
\eq{gyro_poisson}, instead of its Pad\'e approximant, \eq{pade_gyro} (see,
e.g., Grasso \etal ~\cite{grasso_00}).
%
%

For simplicity, let us for now ignore the diffusive and
hyper-diffusive terms and consider an equilibrium described by 
$\bm B_{\perp,eq}=B_0\bm e_y$, with $B_0$ a constant, $n_{e,eq}=0$ and $\phi_{eq}=0$. We want to
linearise \eqs{gyro_ne_1}{gyro_poisson} assuming that all the fields can be
written as $\chi=\chi_{eq}+\chi^1(x,y)$, where $\chi_1$ represents small
perturbations to the equilibrium of the form
$\chi^1(x,y)=\chi_1(x)e^{ik_yy}$. Dropping the subscripts,
\eqs{gyro_ne_1}{gyro_psi_1} become:
\bea
\label{linearized_ne}
\frac{\d n_e}{\d t}&=& f\psi,\\
\label{linearized_psi}
\frac{\d \psi}{\d t}&=& g n_e,
\eea
where
\bea
f&=&-ik_yB_0\kperp^2,\\
g&=&-ik_yB_0\(\ros^2-\frac{\roi^2}{\Gamma_0(b)-1}\).
\eea
Combining these two equations, we have:
\be
\label{linear_0}
\frac{\d^2\psi}{\d t^2}=-\omega^2\psi,
\ee
and similarly for $n_e$, where $\omega^2$ is given by:
\be
\label{KAW_DR}
\omega^2=-fg=\kperp^2 \(\ros^2-\frac{\roi^2}{\Gamma_0(b)-1}\)k_y^2 B_0^2.
\ee
This is the general dispersion relation for the Kinetic Alfv\'en wave (KAW).
In the limit of $\kperp \roi \ll 1$ this relation reduces to:
\be
\omega^2=\[1+\kperp^2 \roi^2\(\frac{3}{4}+\frac{1}{\tau}\)\]k_y^2 B_0^2
\ee
where $\tau=T_i/T_e$. The first term on the RHS constitutes what is usually
referred to as the shear Alfv\'en wave, and the remaining $\kperp^2 \roi^2$
terms are the FLR corrections.  [Note that $|\vec k \cdot \vec B_{eq}|^2 =
k_y^2 B_{eq ,y}^2 = k_y^2 B_0^2$ because $k_z=0$ in this 2-D problem and
the guide field $B_{eq,z}$ only enters through the gyroradii $\rho_i$ and
$\rho_s$.]  The opposite limit of $\kperp \roi \gg 1$ gives:
\be
\label{large_disp_rel}
\omega^2=\kperp^2\rhotau^2 k_y^2B_0^2
\ee
where $\rhotau^2=\ros^2+\roi^2$. The numerical difficulties faced in the
explicit integration of the set of \eqs{gyro_ne_1}{gyro_poisson} at $\kperp
\roi >1$ are obvious from expression (\ref{large_disp_rel}), where we
basically find that $\omega \propto \kperp k_y \propto k^2$.

\subsection{Semi-Implicit operator}
\label{semi-implicit_operator}
Based on the linear dispersion relation, \eq{KAW_DR}, a plausible first choice for the SI operator $\hat \omega^2$ could be:
\be
\hat \omega^2=\kperp^2 \(\ros^2-\frac{\roi^2}{\Gamma_0(b)-1}\)k_y^2 a_0^2 B_0^2,
\ee
where $a_0$ is a dimensionless constant, set to a value that ensures the stability of the algorithm and, to account for a spatially varying equilibrium magnetic field, $B_{eq}=B_{eq}(x)$, $B_0$  can be chosen to be its maximum value.
However, in the nonlinear regime, the magnetic field will no longer be a function of $x$ alone, but rather an arbitrary function of $x$ and $y$. This suggests a qualitative generalization of this expression in order to also apply to the nonlinear regime:
\be
\label{disp_rel_general}
\hat \omega^2=\kperp^4\(\ros^2-\frac{\roi^2}{\Gamma_0(b)-1}\)a_0^2 B_{\perp,max}^2,
\ee
where we simply have made the transformations $k_y\rightarrow\kperp$ and
$B_0\rightarrow B_{\perp,max}$, where we define
$B_{\perp,max}=max(\sqrt{B_{x}^2+B_{y}^2})$ is the maximum magnitude of
total in-plane magnetic field, updated at each new timestep. 
Alternatively, one could try developing an SI operator that included the
spatial dependence of the magnetic field, but it would no longer be
analytically invertible in $k$, and would require alternative solvers
that are not available in our code at present.  (For example, some
rational function approximations for an SI operator that included
spatial variation of $B_0$ might be efficiently invertible in real space
using a multigrid solver.  See Ref. \cite{chacon_03} for some other
possible approaches to preconditioners / SI operators for problems with
strong inhomogeneities, anisotropies, and multiple waves.)
Of course, for problems in which the magnetic field has spatial
variation, as in the application of the next section, \eq{KAW_DR} is no
longer an accurate dispersion relation for the system (i.e., it is no
longer an eigenvalue of the linear problem, and Fourier modes are no
long eigenvectors). 
Nevertheless, as we shall see in \secref{num_tests_sec}, the SI operator
based on a constant $B_0$ is found to be fairly effective for the case
considered there.  (In part this may be because stability of the SI
algorithm only needs the approximation $\hat{\omega}^2$ to be sufficiently
large.  While an SI operator that is larger than necessary may require
more iterations to converge from an arbitrary initial guess, this can be
offset if the initial predictor-corrector step is sufficiently accurate.)
In places where $\kperp\roi\gg1$, this operator simply reduces to the spectral equivalent of the real-space $\nabla_\perp^4$ operator proposed by \cite{harned_mikic_89} for the whistler wave. Note, however, that whereas $\kperp\roi$ will certainly be very large in the vicinity of the X-point, it will also be negligibly small far away. It is therefore very useful to retain the full form of the $\Gamma_0$ operator in the expression for $\hat\omega$. Because our code works in Fourier space, 
this operator is trivial to implement without resorting to any
approximations. Since it does not require any convolution, 
the computational cost of evaluating expression~(\ref{disp_rel_general}) at
each timestep is negligible. In the conclusions we address the issue of how it might be 
generalized to a real-space grid implementation.


An important property of this SI operator is its ability to act on each
value of $\kperp$ individually. For instance, the magnitude of $\hat\omega$
is small for the lower values of $\kperp$, meaning that the large scale
features are left relatively unaffected. Also, it is well behaved in the
limit of $\kperp \roi\ll1$, where it reduces to
\be
\hat\omega=\[1+\kperp^2 \roi^2\(\frac{3}{4}+\frac{1}{\tau}\)\]\kperp^2 a_0^2 B_{\perp,max}^2,
\ee 
a form which makes transparent the operator's ability to equally stabilize the shear Alfv\'en wave.

Finally, we note that the substitution $k_y\rightarrow\kperp$ in
the expression for the SI operator implies that during the linear and early
nonlinear regimes the parameter $a_0$ can be set to values which are
smaller than 1. This is important since, as mentioned, we want the operator
$\hat\omega$ to be sufficiently large to satisfy the stability condition
$\hat\omega^2>\omega^2$ (and ideally, $\hat{\omega}^2$ as close as
possible to $\omega^2$ for all wavenumbers), but not so large that
convergence is slow and the error defined by expression
(\ref{SI_error_cond}) (which is proportional to $a_0^2$) is larger than it
need be.
\subsection{Semi-implicit error control and timestep determination loop}
\label{error_control_sec}
An important part of the algorithm is the semi-implicit error control
and timestep determination loop. The basic idea behind it is to enforce
the limit on the semi-implicit error prescribed by the user, 
\be
\label{error_cond}
\max |\Error_{ji}^p|\leq \Err_max,
\ee
while maximizing the timestep that can be taken. This is done the following way. The discretized set of \eqs{phi_star_final}{phi_final} is iterated until $p=p_{max}$. Then, the following conditions determine the next step:
\bea
\label{next_dt}
\begin{array}{ll}
-~(a)&\Error < 0.8\Err_max : \dt^{n+1}=C^+\dt^n\\
-~(b)&0.8<\Error<\Err_max : \dt^{n+1}=\dt^n\\
-~(c)&\Error>\Err_max : \text{loop is repeated with }\dt^{n+1}=C^-\dt^n
\end{array}
\eea where $C^+>1$ and $C^-<1$ are constants. The goal is to keep the error below but as close to the maximum as possible. Based on the second order convergence of the numerical scheme, we set $C^+=1.08$ and $C^-=0.92$.
At every timestep, the CFL condition for the plasma flow velocities is also evaluated, according to:
\be
\label{CFL_flowcond}
\dt_{CFL}^{flows}=0.1\min{\[\frac{\Delta x}{v_{x,\max}},\frac{\Delta y}{v_{y,\max}}\]}.
\ee
The next timestep is the minimum of expressions (\ref{next_dt}-\ref{CFL_flowcond}).
It is important to note that the error control and timestep determination loop just described makes the choice of $a_0$ less vital to the algorithm than otherwise. The reason is that if the calculation is started with values of $a_0$ and $\dt$ such that the SI scheme is unstable, this will cause the error to diverge until it reaches the maximum allowed, at which point $\dt$ is decreased. This cycle will be repeated until a sufficiently small $\dt$ is found for which the SI scheme is stable at that (fixed) value of $a_0$. Of course, if $a_0$ is set to too small a value, stability of the SI algorithm might only be available at values of $\dt$ much smaller than those which can be achieved at larger $a_0$. 
%
Finally, the reader is again reminded that this error is not the total error of the integration, but simply that introduced by the semi-implicit operator. When the iterative procedure has converged, there is still an underlying error
which is that of the Crank-Nicolson scheme.
\section{Numerical Tests}
\label{num_tests_sec}
As mentioned, we have written a fully parallel pseudo-spectral code to
evolve \eqs{gyro_ne_1}{gyro_poisson}. An explicit time-stepping version of
this code has been in use for some time to study the evolution of the
resistive tearing mode in the single fluid MHD limit
($\roi,\ros=0$). The code uses a Fourier basis set on a 2-dimensional
domain with periodic boundary conditions.  Details and extensive
benchmarking can be found
in~\cite{loureiro_phd}; see also~\cite{loureiro_PRL05} for recent
results. The explicit algorithm uses a modified version of the
Adams-Bashford 3rd-order method that allows for variable
timesteps~\cite{tatsuno_PoP06}, combined with a Crank-Nicolson scheme for
the dissipative terms (in this case, the use of a CN scheme for these terms
is fully justified because $\eta \kperp^2 \dt \sim \eta\ll1$, since
$\dt\sim 1/\omega_{KAW}\sim1/\kperp^2$). The timestep is determined by the
CFL condition, which is evaluated at each timestep according to the formula
\be
\label{CFL_cond}
\dt=0.1\min{\[\frac{\Delta x}{v_{x,\max}},\frac{\Delta y}{v_{y,\max}},\frac{\Delta x}{B_{x,\max}},\frac{\Delta y}{B_{y,\max}},\frac{2}{\omega_{KAW,\max}}\]},
\ee
where $\omega_{KAW,\max}$ is an estimate of the maximum frequency of the KAW on the grid, defined as:
\be
\omega_{KAW,\max}=k_{\perp,\max}\(\ros^2-\frac{\roi^2}{\Gamma_0(b)-1}\)^{1/2}k_{y,\max}B_{\perp,\max}
\ee
The code also allows the resolution in the $y$-direction (i.e., perpendicular to the equilibrium direction) to vary throughout a run, as needed. When more resolution becomes necessary in this direction the run is stopped and the number of grid points doubled. The new data values necessary for this procedure are obtained via linear interpolation between two adjacent points. This is a very useful feature for reconnection runs, as the needs for resolution in the $y$-direction increase greatly from the linear to the nonlinear regime.

The base-case chosen for comparison between the SI method of
\eqs{phi_star_final}{phi_final} and the explicit code is characterized as
follows.
The equilibrium is defined by $\psi^{(0)}=\psi_0/\cosh(x)^2$ and
$\phi^{(0)}=0$. We set $\psi_0=3\sqrt 3/4$~ so that the maximum value of
the equilibrium magnetic field is 1. Other parameters are the resistivity
and the viscosity coefficients, set to $\eta=\nu=5.10^{-4}$, and the
instability parameter $\D'=17.3$ (i.e., the simulation box size is
$L_x=2\pi$, $L_y=2.18\pi$). The ion and  
the ion-sound Larmor radius, $\roi$ and $\ros$, are set to 0.02. 
The hyper-resistivity $\eta_H$ and hyper-viscosity $\nu_H$ are grid dependent, and their value is calculated at each timestep, 
according to the formula:
\be
\eta_H=\nu_H=0.1~\omega_{KAW,\max}/k_{\perp, \max}^4.
\ee 
Tests have been performed to insure that these coefficients are
sufficiently small not to alter the physics of the system. The tearing mode
instability is initialized by perturbing this equilibrium with
$\psi^{(1)}=-10^{-5}\cos(2\pi/L_y~y)$. The initial resolution is
$3072\times128$ grid points; the number of grid points the $y$-direction is
doubled throughout the run, up to 2048 for $t>\approx210$.
\begin{figure}
\begin{center}
\epsfig{file=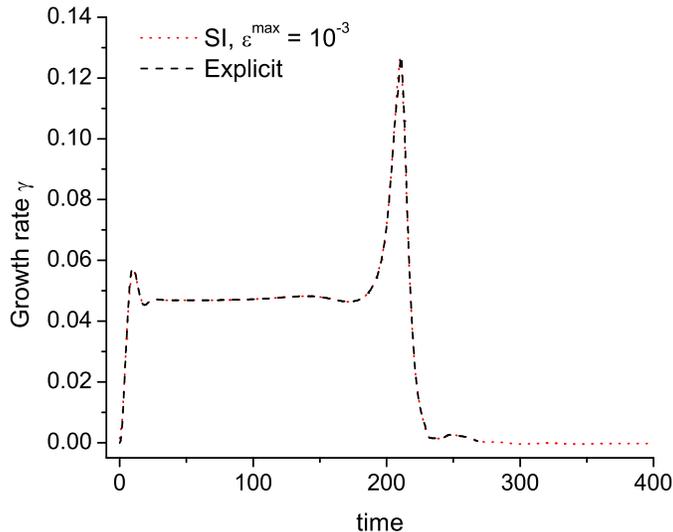,width=10cm} 
\end{center}
\caption{\footnotesize{Comparison between the explicit and SI results. The
Figure shows the growth rate of the tearing instability, $\gamma=d \log
\psi_X/ dt$, obtained by the explicit method (dashed black line) and
the SI method with $\Err_max=10^{-3}$ (dotted red line). The two curves
show excellent agreement and visually overlap exactly.	}}
\vspace{0.5cm}
\label{grate_comparison}
\end{figure}
\begin{figure}[htp!]
\unitlength1cm
\begin{center}
	\begin{tabular}{cc}
		\epsfig{file=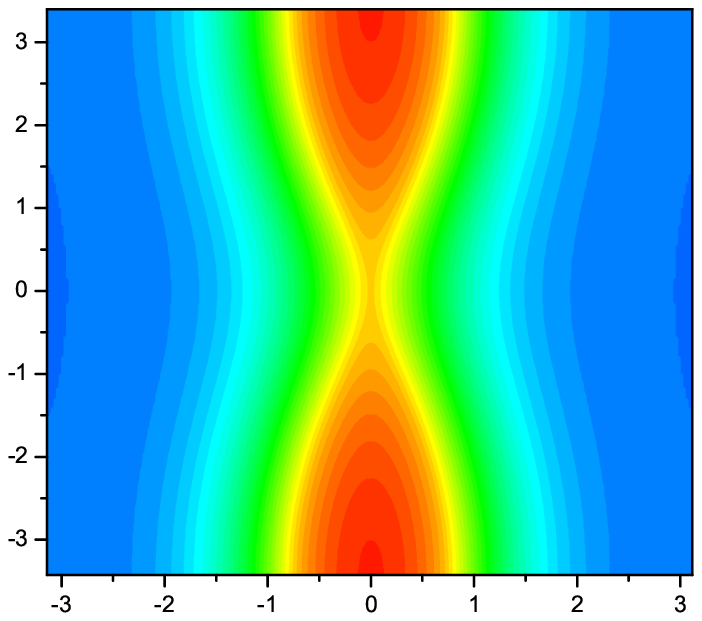,width=6.8cm} &
		\epsfig{file=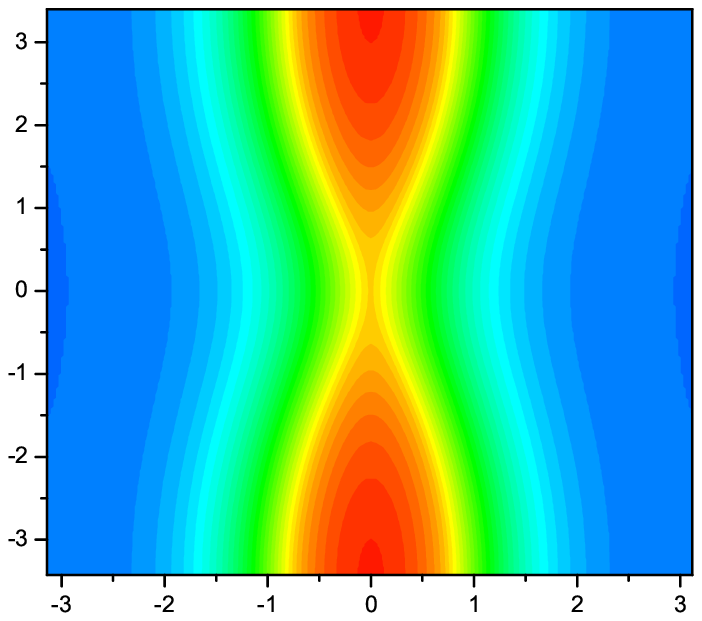,width=6.8cm} \\
		\epsfig{file=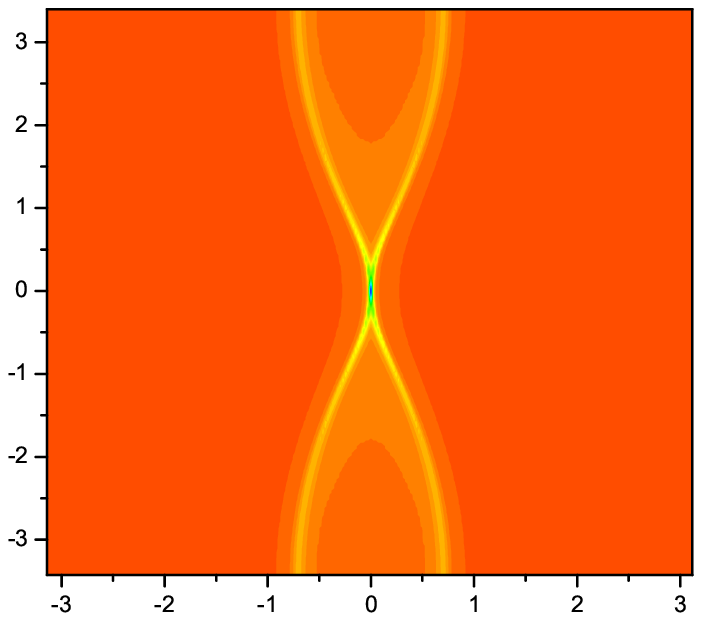,width=6.8cm} &
		\epsfig{file=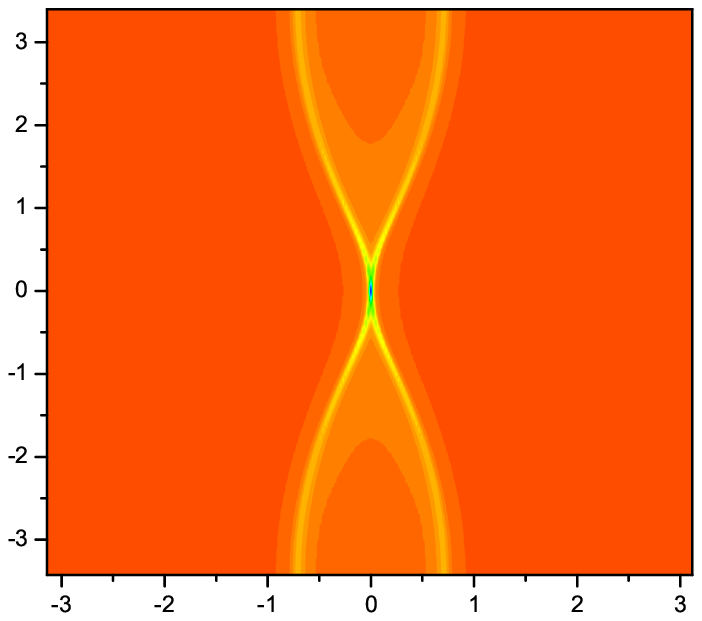,width=6.8cm} \\
		\epsfig{file=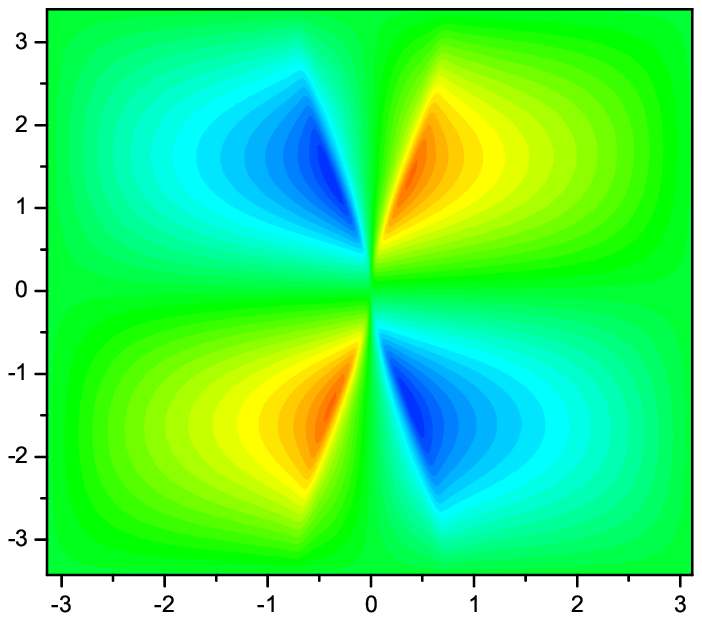,width=6.8cm} &
		\epsfig{file=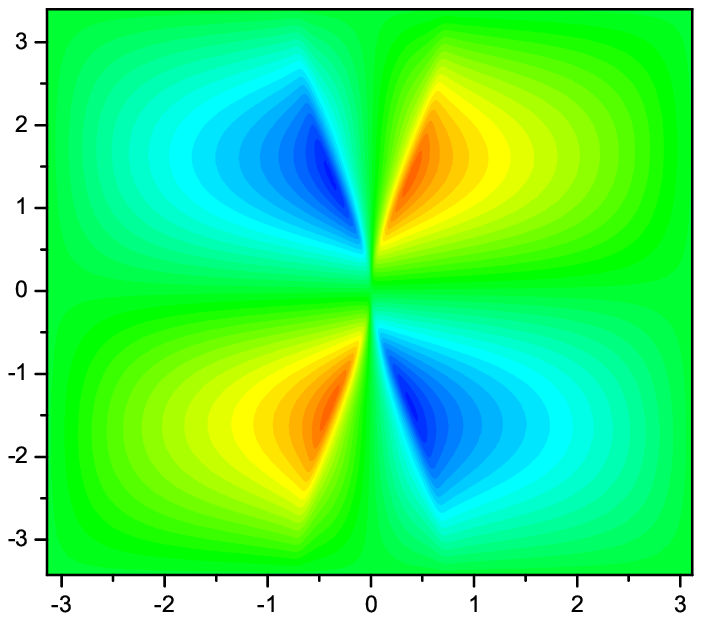,width=6.8cm} \\
		\epsfig{file=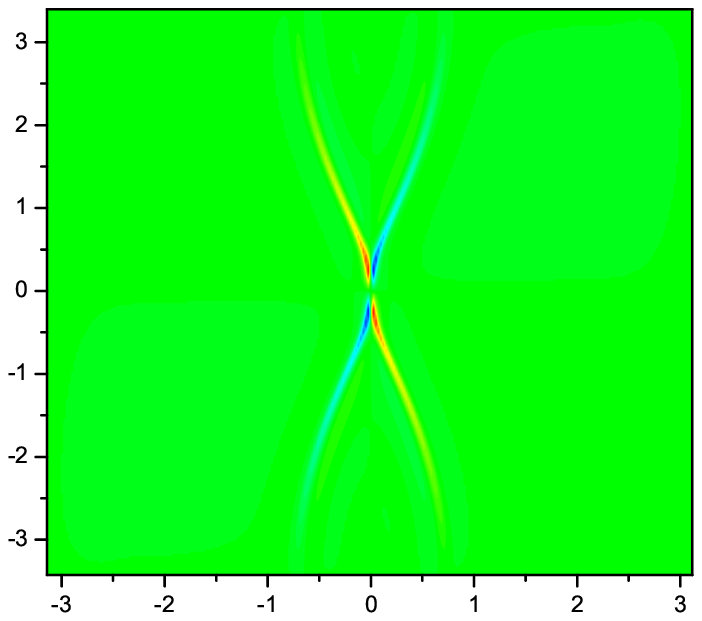,width=6.8cm} &
		\epsfig{file=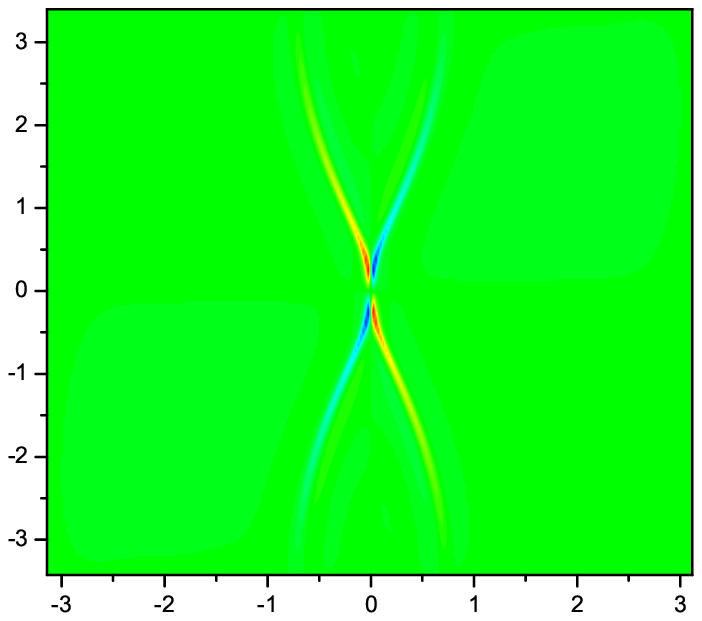,width=6.8cm} \\
		\end{tabular}
\end{center}
\caption{\footnotesize{Contour plots of the fields measured at an island width $W\approx1.6$ ($t\approx204$). Left column show results obtained with the explicit integration scheme, right column those obtained with the semi-implicit scheme. From top to bottom, shown are the contours of $\psi, j_{\parallel}, \phi$ and $n_e$.}}
	\label{expl_vs_SI}
\end{figure}
\begin{figure}[ht!]
   \begin{center}
	\begin{tabular}{c}
		\epsfig{file=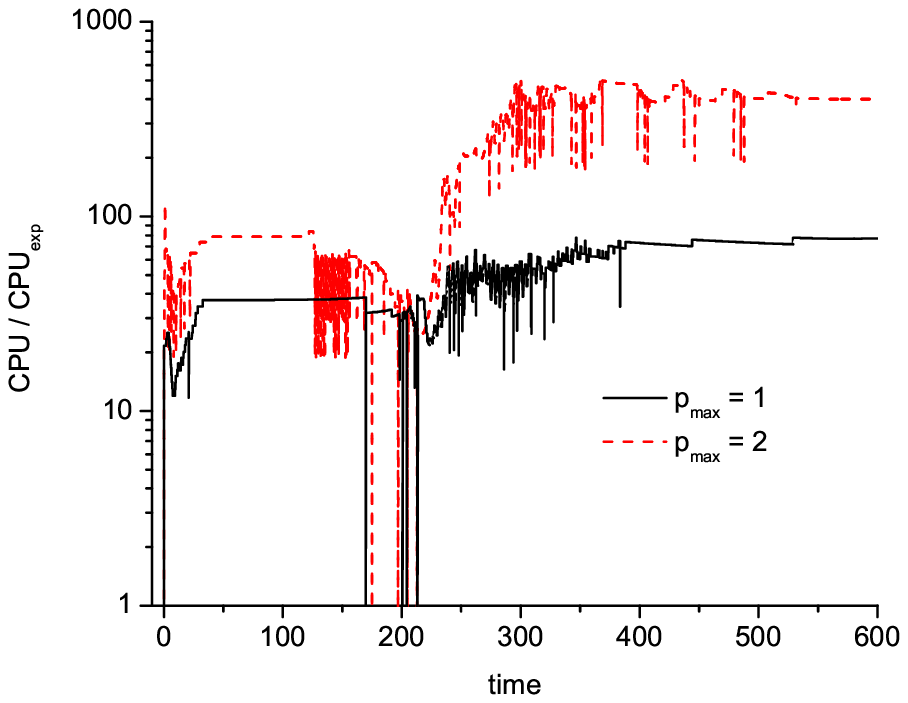,width=10cm} \\
		(a)	\\
		\epsfig{file=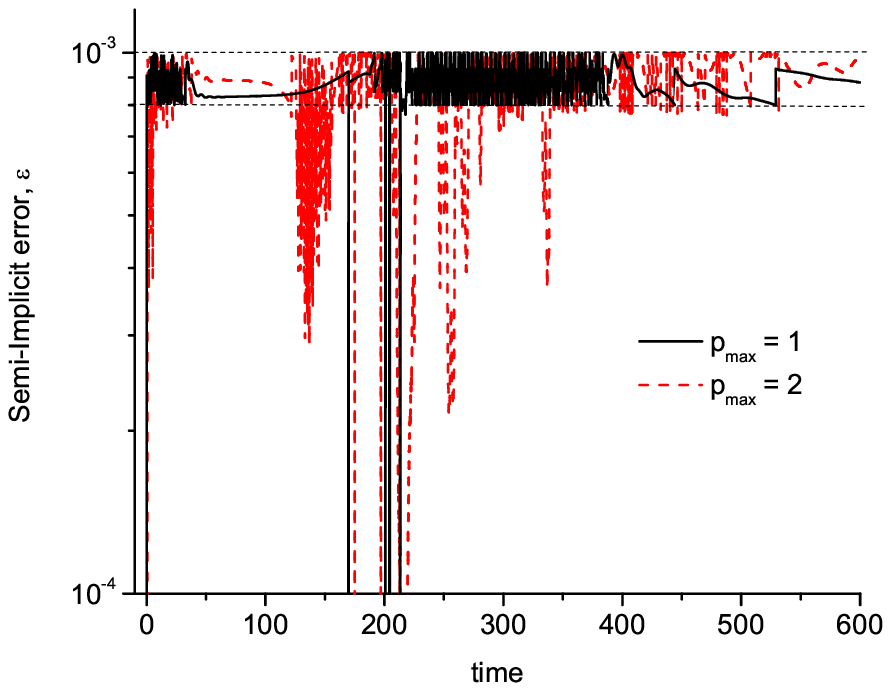,width=10cm} \\
		 (b)
	\end{tabular}
   \end{center}
	\caption{\footnotesize{Fig.(a) shows the CPU speed-up factor over a conventional explicit integration for $p_{max}=1,2$. The timestep taken by the SI scheme is determined by imposing $\mathcal E <10^{-3}$, as seen in Fig.(b). The dashed horizontal lines limit the error ``comfort zone'', defined by expression~(\ref{next_dt}). The largest undershoots in both figures (at times in the vicinity of $t\approx200$) are due to code restarts at those times, when the resolution in the $y$-direction is doubled. }}
	\label{dt_ratio}
\end{figure}
The most crucial test is to check how closely the semi-implicit method described in this paper reproduces the results obtained by a conventional explicit integration. This comparison is performed in \figs{grate_comparison}{expl_vs_SI}. \fig{grate_comparison} shows the growth rate, defined as $\gamma=d \log \psi_X/ dt$, where $X$ is the location of the $X$-point, obtained by the two methods. The agreement between both approaches is remarkable. After an initial transient ($t<\sim30$), the tearing instability is seen here to evolve through four different stages: the linear stage (constant growth rate), the early nonlinear period, from $t\approx130 - 170$, followed by the explosive stage, during which the growth rate dramatically increases ($t\approx170 - 220$) and finally the growth rate slow down and saturation period.

A comparison between the contour plots for all the fields at the early
stages of the explosive period  ($t=204$) is shown in
\fig{expl_vs_SI}. Again, the figure yields excellent agreement between the
explicit and the semi-implicit approaches.

\begin{figure}
	\centering
		\epsfig{file=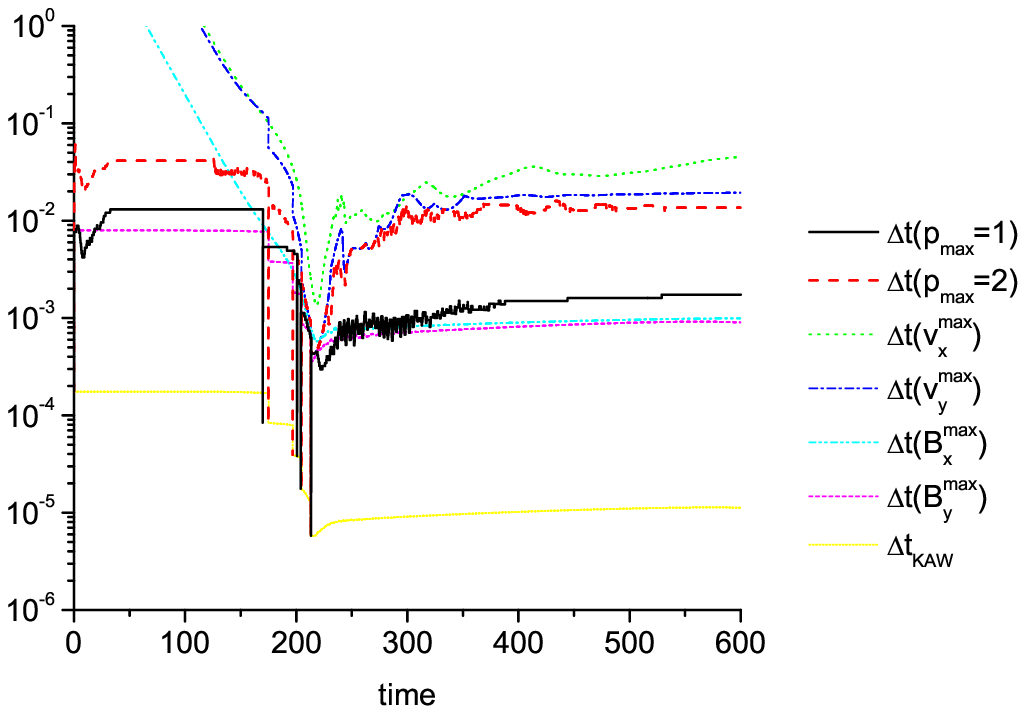,width=13cm}
	\caption{\footnotesize{Plot of $\dt$ \textit{vs.} time achieved by the numerical scheme of \eqs{phi_star_final}{phi_final} for the test case and for $p_{max}=1$ (black full line) and $p_{max}=2$ (red dashed line). Also plotted for comparison are the timesteps due to each term contributing to the CFL condition (\ref{CFL_cond}).  Note that the timestep taken by the semi-implicit method coincides with that based on the flow velocity during the explosive part of the instability.}}
	\label{dt_vs_flows}
	\vspace{0.5cm}
\end{figure}

Plotted in \fig{dt_ratio}(a) is the CPU speed-up factor, determined as follows:
\be
\frac{CPU}{CPU_{exp}}=\frac{\dt}{(N_{RHS}/2) \dt_{exp}}
\ee
where $N_{RHS}$ is the number of evaluations of the RHS required at each timestep for the SI scheme to converge, compared to an explicit scheme that requires only 2 such evaluations per timestep. $CPU_{exp}$ and $\dt_{exp}$ are the explicit integration CPU time and timestep, respectively. As mentioned, convergence is determined by $\mathcal E \le 10^{-3}$ at each timestep. A plot of the time evolution of the maximum error is shown in \fig{dt_ratio}(b). 
The dashed lines in this figure identify the ``comfort zone'' for the SI error, defined by expression (\ref{next_dt}). We see that for both values of $p_{max}$ plotted, the error is indeed largely confined to that region, indicating the second-order convergence of the iterative scheme.
The large undershoots mark the points at which the run was restarted with a larger resolution in the $y$-direction. At each restart, the $\dt$ is set to be equal to that determined by the CFL condition. 
As shown in~\fig{dt_ratio}(a), the SI method is remarkably successful, yielding a CPU speed-up factor ranging from a minimum of $\sim20$ during the explosive growth period, to several hundreds during nonlinear saturation of the instability. We also observe that, in this case, the iterative scheme is very efficient, as is shown by the larger speed-up factors allowed for $p_{max}=2$ (red dashed line). 
The actual timesteps taken by the SI method are shown in \fig{dt_vs_flows}. We see that the SI method performs so well that the minimum CPU enhancement ($\sim20$ at $t\approx210$) is set by the flow speed, i.e., the actual dynamics that we are interested in following (recall that the CFL condition for the flows is imposed as a maximum for the timestep). 
\subsection{Numerical test of the second-order accuracy of the algorithm}
We discussed the 2cd-order nature of the algorithm in \secref{num_test_sec}, 
and in \secref{sec_ord_op_split} we provided an analytical demonstration that
the algorithm remains second-order accurate for sufficiently small $\Delta
t$ even if the diffusion and advection operators do not commute.  These
arguments are rigorously valid only if $\dt \d \FF/\d \psi \ll 1$.
However, the goal of implicit and semi-implicit algorithms is to allow
usage of very large time steps compared to explicit algorithms.  Thus,
although the dominant modes of interest have eigenfrequencies
$\omega$ of $i \d \FF / d \psi$ such that $\dt \omega$ is presumably small,
there will be some high frequency modes for which the effective $\dt \d
\FF/\d \psi$ is large.
To demonstrate that the scheme can indeed be second-order accurate (at
least for our test case) at desirable implicit values of 
$\dt$ (i.e., which greatly exceed those required by an explicit 
integration), 
we performed a scaling study of the error of the algorithm with the 
timestep $\dt$, using as a test case the application discussed in 
the preceding section.
This study was done as follows: starting at time $t\approx 210$, when
the simulation is fully nonlinear (see~\fig{grate_comparison}) and the
SI algorithm is using a time step of $\dt = 9 \times 10^{-4}$, we run the
code for an additional time of $0.135$ using a very small time step of
$\dt=10^{-6}$.   This value is smaller than even the explicit timestep 
$\dt \sim 1.5 \times 10^{-5}$ that would be required at this stage, and
we take the result of this  integration as our exact result. Then, we
perform several runs over this same time period of 0.135 with various
fixed values of the time step $\dt$.
To isolate the scaling of the error with $\dt$,
this scaling study was performed with a fixed number of iterations per
time step, $\pmax=2$.
The error is the difference between   
the results of these integrations and the ``exact'' result 
(we have chosen to plot the relative error in $\psi$ at the $X$-point).
\begin{figure}[tbp]
	\centering
 		\includegraphics{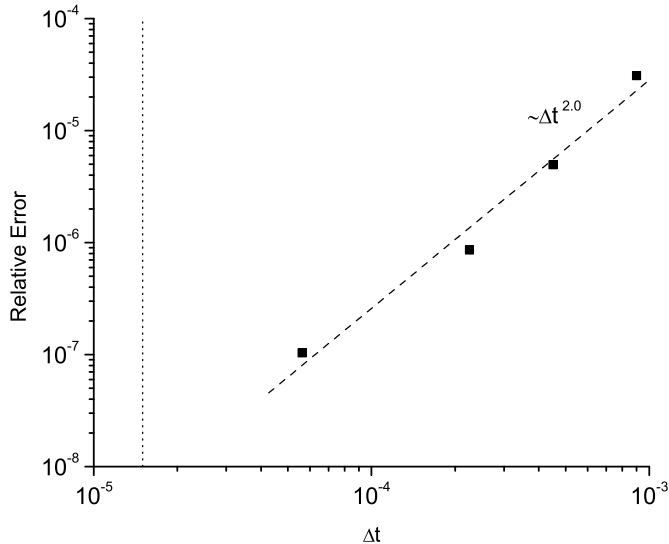}
 	\caption{Scaling of the relative error with $\dt$.  Dashed line
 	indicates $(\dt)^2$ slope.  Vertical dotted line marks the explicit
 	timestep limit that would be required at this point in the simulation.
 	\label{error_scaling}}
\end{figure}
The result is shown in \fig{error_scaling}, where a clear scaling of the error with 
$(\dt)^2$ is seen.
Note that, at each value of the iteration index $p$, we have $\psi^{n,p}=\psi_{true}+\mathcal E_{SI}^{n,p}+\mathcal E_{discr.}^{n}$ (and similarly for $\phi$), where $\psi_{true}$ is the true solution 
(here taken to be that given by the integration with $\dt=10^{-6}$),
$\mathcal E_{SI}$ is the error introduced by
the SI operator, and $\mathcal E_{discr.}$ is the error due to the 
particular discretization scheme used (e.g., when our scheme is converged 
in the $p$-iterations, $\mathcal E_{discr.}$ is simply the error of the 
CN method).   
Thus what is plotted in the figure is the total error, due to both the SI
error and the time discretization error.
In the figure, the vertical dotted line identifies the timestep 
that would have been required for stability by an explicit algorithm at
this point of the simulation.  
%
%
The dashed line shows a $\dt^2$ slope. 
We thus see that the second-order accuracy of the 
algorithm extends at least up to the typical SI values of $\dt \sim 9
\times 10^{-4}$, which is 60 times larger than the explicit time step limit.
\section {Conclusions}
\label{conclusions}
We have presented an efficient and accurate numerical method for the time
integration of stiff wave coupled partial differential equations. The
method allows for timesteps which greatly exceed those imposed by the CFL
constraint, while accurately reproducing explicit calculations.
CPU time savings over a conventional explicit scheme range
from factors of $\sim20$ to several hundreds for the test case chosen
(gyrofluid magnetic reconnection).
The method is second-order accurate even if the diffusion and nonlinear
(advection) operators do not commute, and exhibits the property of robust
damping.  I.e., it correctly captures the analytic solution in the limit
$\etak\dt\gg1$, where $\etak$ is the diffusion operator and $\dt$ the
timestep.  In the limit of no physical dissipation, the method
is symplectic (at least for a linear wave test problem), i.e., it does not
introduce any numerical dissipation.

The derivation of the SI numerical method presented here relies on
the physics based assumption that the fundamental wave dynamics arises as a
result of coupling between equations. In the linear regime (small
perturbation amplitude) this statement is exact for many physical
systems. Our results demonstrate that this same SI operator (scaled to
approximate an upper bound on the frequency in the nonlinear regime)
continues to be effective in the nonlinear regime as well, at least for the
test problem explored here.  
We note that the application chosen to study the efficiency of our algorithm
supports only one wave (the KAW).  It is therefore easier to formulate
an efficient SI operator (or an efficient preconditioner for an implicit
algorithm) for this system that would be the case for a system of
equations that supported multiple waves (i.e., multiple polarizations at
each wave number).  While the isotropic and homogeneous SI operator used
here was fairly effective for the test case we considered, which had
some degree of anisotropy and inhomogeneity, there may be cases with
stronger anisotropies or inhomogeneities where a more complicated SI
operator or preconditioner is needed.  Some approaches for dealing with
these issues (of multiple waves with anisotropies and inhomogeneities)
are discussed in Ref. \cite{chacon_03}.  The efficiency of the iterative
SI algorithm for a multiple wave system remains to be tested.

One of the advantages of a physics-based
semi-implicit operator is that one only has to approximate $\omega^2(\vec
k)$, not the dispersion relation $\omega(\vec k)$ itself (and in fact, one
only needs an upper bound on $\omega^2$ to insure convergence).  Chac\'on and
Knoll \cite{chacon_03,knoll_keyes_04,knoll_05} have discussed how this
kind of physics-based semi-implicit operator can be related to
preconditioning, and how it turns an originally hyperbolic problem into a
diagonally dominant parabolic problem that can be efficiently solved by
multigrid and/or preconditioned Krylov (if the preconditioner is
sufficiently effective) iterative techniques.  In some cases, the resulting
parabolic problem can also be efficiently solved with FFTs, which we use
here.

The Iterative SI approach used here provides a way to monitor and control
the SI operator splitting errors, and essentially turns a semi-implicit
algorithm into a fully implicit algorithm.  Accuracy, rather than
stability, sets the time step $\dt$.  There are several variations of SI
algorithms in the literature (such as
\cite{harned_kerner_85,harned_86,schnack_87,harned_mikic_89,lionello99,caramana_91}),
some using staggered time grids.  If only one $\pmax=1$ iteration is done, the
underlying SI algorithm used here is equivalent to reference
\cite{harned_86}.  The iterative approach presented here could also be
applied to other variations of SI algorithms.  This could be a
straightforward way to extend other existing codes that use SI algorithms.

It is pertinent to compare the numerical method developed here to more
elaborate implicit schemes and preconditioning strategies. Within these,
Jacobian-free Newton-Krylov (JFNK) methods seem to be amongst the most
sophisticated and promising~\cite{knoll_keyes_04}.  (As discussed in
\secref{equiv_precond}, our algorithm has similarities to a Jacobian-Free
Newton method, but with simple functional iteration instead of Krylov iteration
within each Newton step.)  In this respect, the main comparison factor is
the CPU time speed-up over explicit approaches -- the main purpose of
implicit schemes is, after all, to maximize this parameter. Of course, this
is highly problem specific, and a general comparison is not possible. For a
similar problem to the one dealt with in this paper, i.e., implicit
integration of high frequency dispersive plasma waves, a JFNK implicit
scheme was developed by Chac\'on and Knoll~\cite{chacon_03}. The aim of
these authors is the accurate implicit integration of the whistler wave,
which has a dispersion relation of the same type as the KAW, i.e.,
$\omega\sim\kperp^2$. In that work, CPU time enhancements by factors of up
to a maximum of $\sim 30$ is quoted.  This is roughly comparable to the
enhancement factors of 30 and higher found for our method, as shown in
\fig{dt_ratio}(a).  However, a definitive comparison is not warranted at
this time, since both the physics model and the problem are different
(Chac\'on and Knoll use a more complex four-field model, the flux-bundle
coalescence problem is different than the reconnection problem done here,
grid sizes differ, there are subtle differences in the definition of the
CFL condition that influence this comparison, and the amount of speedup
depends sensitively on the stage of the nonlinear dynamics in our
simulation). 

It is interesting to note that the case in Table I of \cite{chacon_03} that
shows the largest speedup factors also corresponds to needing no Krylov
iterations for each Newton iteration (as the authors point out, the preconditioner
itself is providing a very good initial guess). In that case, the algorithm
of reference~\cite{chacon_03} becomes similar to the one presented here, 
(except for differences
in the predictor step and in the treatment of dissipation terms), so one
would in fact expect comparable performance.  But short of more detailed
comparisons, one should simply conclude that the Iterative Semi-Implicit
method developed here seems competitive, at least for some problems, with
these somewhat more complex approaches.  The Iterative Semi-Implicit
algorithm has the relative advantage of being simpler to implement
(particularly for codes that are already semi-implicit) and works well for
the type of problem we have studied here, while a full JFNK algorithm would
be more robustly convergent and may be more efficient for other types
of problems where the present algorithm needs more than a few iterations. 

With regards to the implementation of a similar treatment of 
the diffusive terms in real space codes, we begin by noting that, in the final formulation of the numerical scheme \eqs{phi_star_final}{phi_final}, what has to be evaluated are
operators like $e^{-\DD \Delta t}$.  One way to evaluate this with a
real-space code could be with a variation of Exponential Propagation
Iterative methods~\cite{tokman06}. Another way would be to use a
2cd-order rational-function approximation
\be
e^{-\DD \Delta t} \approx 
   (1 + a \DD \Delta t)
   (1 + b \DD \Delta t)^{-1} (1 + b \DD \Delta t)^{-1}, 
\nonumber
\ee
with $a=1+\sqrt{2}$, $b=(2+\sqrt{2})/2$.  This retains
the second order accuracy of a Crank-Nicolson treatment of
damping while having the advantage of robust damping in the large
$\DD \Delta t$ limit (unlike Crank-Nicolson where $e^{-\DD \Delta t}
\approx (1 + \DD \Delta t/2)^{-1} (1 - \DD \Delta t/2) \rightarrow
-1$), and still remaining relatively easy to evaluate, since inverting the
$(1+b \DD \Delta t)$ operator is equivalent to a standard implicit step. Note, however, that this has not been tested by 
us in this work.

Our code works in Fourier-space, where it is very easy to invert the SI
operator presented in \secref{semi-implicit_operator}, but it should 
also be possible to implement a very similar SI operator
in real space.  Using a Pad\'e approximation for the $\Gamma_0$ function
and Fourier-transforming to real space, \eq{disp_rel_general} becomes $\hat
\omega^2= - a_0^2 B_{\perp,max}^2 \nabla_\perp^2 \(1 - \rho_\tau^2
\nabla_\perp^2 \)$.  The resulting SI operator $(1 + \dt^2 \hat{\omega}^2
/4)$ is a symmetric positive-definite parabolic operator that could be
solved with a preconditioned iterative Krylov solver or multigrid
algorithm.  (In some cases, a direct sparse-matrix solver can also be
used.) It should be noted that if a Krylov solver is used, it
is important to use a very good preconditioner.  For these types of
parabolic problems without preconditioning, the number of iterations
required for Krylov solvers like Conjugate Gradients or GMRES scales with
the square root of the condition number of the
matrix~\cite{Saad96,Axelson96}, $N_{iters} \propto \sqrt{1 + \dt^2
\hat{\omega}_{max}^2/4} \propto \dt \hat{\omega}_{max}$, which would give
it at most an order unity speed up over an explicit code.  To get a
significant speed up, an iterative solver must be aided by further
algebraic preconditioning, such as Modified-ILU~\cite{gustafsson78},
multigrid, or multilevel additive Schwarz~\cite{knoll_keyes_04}, or some
combination thereof. 
Again, we note that this has not been tested by us: in our application, 
direct inversion of 
the operator (\ref{disp_rel_general}) is trivial since we use a pseudo-spectral code.

An important observation from the physics point of view is that these results, 
i.e., the fact that the KAW can be integrated implicitly without affecting the 
physics of the system, confirm the parasitic role of this wave in the reconnection 
process, although the terms from which they arise are essential to the observed 
speed up of the instability growth rate. A similar conclusion has been reported 
by Chac\'on and Knoll~\cite{chacon_03} for the whistler wave.

Finally, we note that this numerical method could be useful for
other problems where an implicit treatment of high-frequency waves is
desired, and in particular is trivially applicable to fluid studies of
collisionless reconnection, where electron inertia is added to
\eqs{gyro_ne_1}{gyro_poisson} and replaces the role played by the
resistivity in breaking the frozen flux condition.

\section*{Acknowledgments}

It is a pleasure to acknowledge a very useful discussion with L. Chac\'on.
Conversations with R.~Samtaney and V.~S.~Lukin are also gratefully
acknowledged. This work was funded by The Center for Multiscale Plasma
Dynamics, the U.S. Dept. of Energy Grant No. DE-FC02-04ER54784, and at
the Princeton Plasma Physics Laboratory by U.S. Dept. of Energy Grant
No. DE-AC02-76CH03073.
\appendix
\section{Exact integration of the diffusive exponential in \eqs{step1_psi}{step1_phi}}
\label{exact_exp}
It is instructive to consider one logical alternative to the discretization
of \eqs{step1_psi}{step1_phi} performed in \eqs{step2_psi}{step2_phi}. For
simplicity, let us instead consider the more compact form of the problem
given by \eq{compact_eq2}.
A formal solution to this equation can be written as:
\be
\tilde \psi^{n+1}=\tilde \psi^n+\int_{t_n}^{t_{n+1}}dt~e^{\etak t}\FF(\psi).
\ee
In the Crank-Nicolson algorithm of \eqs{step2_psi}{step2_phi}, what is
done is to approximate the integral with a 2cd order midpoint evaluation,
i.e.:
\be
\int_{t_n}^{t_{n+1}}dt~e^{\etak t}\FF(\psi)\approx\frac{\dt}{2}\[ e^{\etak t_n}\FF(\psi^n)+e^{\etak t_{n+1}}\FF(\psi^{n+1})\],
\ee
from which formulas of the kind of (\ref{step2_psi},\ref{step2_phi}) directly follow.
Instead, one can exactly integrate the exponential term, while using the CN discretization for the $\FF$ operator alone. Undoing the variable substitutions of (\ref{varsubs}), we obtain:
\be
\psi^{n+1}=e^{-\etak \dt}\psi^n+\frac{\dt}{2}\(\frac{1-e^{-\etak
    \dt}}{\etak \dt} \)
\[\FF(\psi^n)+\FF(\psi^{n+1}) \].
\ee
Now following a procedure similar to that of \secref{sec_ord_op_split}, it can be shown that this method 
reduces to \eq{sec_order} in the limit of $\FF\dt\ll1$, $\etak\dt\ll1$, i.e., it
is also second-order accurate even if the operators $\FF$ and $\etak$ do
not commute. It does not, however, exhibit the property of robust damping
discussed in \secref{robust_damp} as can easily be seen by evaluating the
above expression in the limit $\etak\dt\gg1$ (but it does correctly treat 
the case of $\FF(\psi)=const.$, which our proposed algorithm does not, 
see appendix B. A method to accurately deal with both cases in the 
limit of $\etak\dt\gg1$ is left to a future publication).

Yet another second-order accurate alternative is
Strang-splitting~\cite{strang_68}.  For the model problem \eq{robust_eq},
one form of Strang-splitting is $\psi^{n+1} = e^{-\etak \dt/2} (1+i \omega 
\dt/2)^{-1} (1 - i \omega \dt /2) e^{-\etak \dt/2} \psi^{n}$.
Similarly to our numerical scheme of \eqs{phi_star_final}{phi_final}, this
also exhibits robust-damping. However, we found this method to be much less
efficient in terms of the timestep enhancements that can be obtained over
the CFL condition. Further tests would be required to understand why our
splitting scheme is a better alternative than Strang-splitting for this
problem; the interested reader is referred to a recent paper by Kozlov
$\etal$~\cite{kozlov_04} on the nature of the local error for splitting
methods applied to stiff problems.
\section{Inclusion of the equilibrium source term in \eqs{phi_star_final}{phi_final}}
\label{eq_term}
The equilibrium source term on the RHS of \eq{gyro_psi_1}, $E_{ext}=\etak\psi_{eq}$, is not explicitly accounted for in the derivation of the numerical method of \eqs{phi_star_final}{phi_final}. Since $\FF$ is a general operator, it can, of course, include this term. A more accurate alternative is obtained using the fact that $\d \psi_{eq}/\d t=0$. Thus, \eq{gyro_psi_1} can be recast in terms of $\psi_1=\psi-\psi_{eq}$ in the form
\be
\frac{\d \psi_1}{\d t}=\FF(\phi,n_e,\psi)-\etak\psi_1.
\ee
Now defining $\psi_1=e^{-\etak t}\tilde\psi_1$ we obtain
\be
\frac{\d \tilde\psi_1}{\d t}=e^{\etak t}\FF(\phi,n_e,\psi),
\ee
which is of the same form as \eq{step1_psi} and so the derivation of the method can be carried out as in \secref{derivation_method}. \Eqs{psi_star_final}{psi_final} are thus replaced by:
\bea
\label{psi_star_final_eq}
\psi^{n+1,*}&=&e^{-\etak\dt}\psi^n+\(1-e^{-\etak\dt} \) \psi_{eq} \nonumber\\
& &+\frac{\dt}{2}\(1+e^{-\etak \dt}\)\FF(\phi^n,\psi^n),\\
\label{psi_final_eq}
\psi^{n+1,p+1}&=& e^{-\etak \dt}\psi^n +\(1-e^{-\etak\dt} \) \psi_{eq}
+ \frac{\dt}{2}e^{-\etak \dt}\FF(\phi^n,\psi^n)+\nonumber\\
& &+\frac{\dt}{2}\FF(\phi^{n+1,p},\psi^{n+1,p})
-\frac{\hat\omega^2\dt^2}{4}(\psi^{n+1,p+1}-\bar\psi^{n+1,p}).
\eea


\end{document}